\documentclass[reprint,floatfix,twocolumn,superscriptaddress,showpacs, aps, prx]{revtex4-2} 

\usepackage{graphicx} 
\usepackage{natbib} 
\usepackage[usenames,dvipsnames]{color} 
\usepackage{amsmath} 
\usepackage[urlcolor=blue, hyperindex, colorlinks, bookmarks=true,linkcolor=black,citecolor=black]{hyperref} 
\usepackage{dcolumn}
\usepackage{amssymb} 
\usepackage{soul} 
\usepackage{ifthen} 
\usepackage{bbm}
\usepackage{comment}
\usepackage[ampersand]{easylist}
\usepackage{upgreek}

\def\l{\left}
\def\r{\right}

\def\shownoteal{1} 
\newcommand{\nal}[1]{\ifthenelse{\shownoteal=1}{\textcolor{red}{[[#1]]}}{}}
\def\shownoteay{1} 
\newcommand{\nay}[1]{\ifthenelse{\shownoteay=1}{\textcolor{orange}{[[#1]]}}{}}

\begin{document}

\title{Calibration of flux crosstalk in large-scale flux-tunable superconducting quantum circuits}

\author{X. Dai}\thanks{x35dai@uwaterloo.ca}
\affiliation{Institute for Quantum Computing, and Department of Physics and Astronomy,
University of Waterloo, Waterloo, ON, Canada N2L 3G1}
\author{D. M. Tennant}
\altaffiliation{Current address: Lawrence Livermore National Laboratory, Livermore, California 94550, USA}
\author{R. Trappen}
\affiliation{Institute for Quantum Computing, and Department of Physics and Astronomy, University of Waterloo, Waterloo, ON, Canada N2L 3G1}
\author{A. J. Martinez}
\affiliation{Institute for Quantum Computing, and Department of Physics and Astronomy, University of Waterloo, Waterloo, ON, Canada N2L 3G1}
\author{D. Melanson}
\altaffiliation{
Current address: Department of Engineering Physics, McMaster University, 1280 Main St., Hamilton, Ontario, Canada L8S 4L8
}
\affiliation{Institute for Quantum Computing, and Department of Physics and Astronomy, University of Waterloo, Waterloo, ON, Canada N2L 3G1}
\author{M. A. Yurtalan}
\affiliation{Institute for Quantum Computing, and Department of Physics and Astronomy, University of Waterloo, Waterloo, ON, Canada N2L 3G1}
\affiliation{Department of Electrical and Computer Engineering,
University of Waterloo, Waterloo, ON, Canada N2L 3G1}
\author{Y. Tang}
\altaffiliation{Current address: 1QB Information Technologies (1QBit), Vancouver, BC, Canada
V6E 4B1}
\affiliation{Institute for Quantum Computing, and Department of Physics and Astronomy, University of Waterloo, Waterloo, ON, Canada N2L 3G1}
\author{S. Novikov}
\affiliation{Northrop Grumman Corporation, Linthicum, Maryland 21090, USA}
\author{J. A. Grover}
\affiliation{Northrop Grumman Corporation, Linthicum, Maryland 21090, USA}
\author{S. M. Disseler}
\affiliation{Northrop Grumman Corporation, Linthicum, Maryland 21090, USA}
\author{J. I. Basham}
\affiliation{Northrop Grumman Corporation, Linthicum, Maryland 21090, USA}
\author{R. Das}
\author{D. K. Kim}
\author{A. J. Melville}
\author{B. M. Niedzielski}
\author{S. J. Weber}
\author{J. L. Yoder}
\affiliation{MIT Lincoln Laboratory, 244 Wood Street, Lexington, Massachusetts 02421, USA}
\author{D. A. Lidar}
\affiliation{Departments of Electrical \& Computer Engineering, Chemistry, and Physics, and Center for Quantum Information Science \& Technology, University of Southern California, Los Angeles, California 90089, USA}
\author{A. Lupascu}\thanks{adrian.lupascu@uwaterloo.ca}
\affiliation{Institute for Quantum Computing, and Department of Physics and Astronomy,
University of Waterloo, Waterloo, ON, Canada N2L 3G1}
\affiliation{Waterloo Institute for Nanotechnology, University of Waterloo, Waterloo, ON, Canada N2L 3G1}

\date{ \today}

\begin{abstract}
Magnetic flux tunability is an essential feature in most approaches to quantum computing based on superconducting qubits. Independent control of the fluxes in multiple loops is hampered by crosstalk. Calibrating flux crosstalk becomes a challenging task when the circuit elements interact strongly. We present a novel approach to flux crosstalk calibration, which is circuit model independent and relies on an iterative process to gradually improve calibration accuracy. This method allows us to reduce errors due to the inductive coupling between loops. The calibration procedure is automated and implemented on devices consisting of tunable flux qubits and couplers with up to 27 control loops. We devise a method to characterize the calibration error, which is used to show that the errors of the measured crosstalk coefficients are all below $0.17$\%.
\end{abstract}

\maketitle

\section{Introduction}
Many prominent quantum computing platforms rely on electrical controls for qubit manipulation and readout~\cite{kjaergaardSuperconductingQubitsCurrent2020, vandersypenInterfacingSpinQubits2017,romaszkoEngineeringMicrofabricatedIon2020,childressDiamondNVCenters2013}. Crosstalk is an important control issue and a major challenge to scaling up quantum computers, in particular for superconducting and semiconducting qubits~\cite{kellyPhysicalQubitCalibration2018,millsComputerautomatedTuningProcedures2019,nowackSingleShotCorrelationsTwoQubit2011}. For superconducting qubits, crosstalk includes control crosstalk, such as crosstalk between DC-bias lines~\cite{abramsMethodsMeasuringMagnetic2019} or microwave control lines~\cite{ rigettiFullyMicrowavetunableUniversal2010,ficheuxFastLogicSlow2020}, and parasitic coupling between qubits ~\cite{chowSimpleAllMicrowaveEntangling2011, albash_2015_consistencytestsclassical}.

Flux control is an important resource for superconducting-qubit-based quantum computers. For gate-model quantum computing implementations, flux tunability is used to individually control the transition frequency of qubits to compensate for fabrication imprecision, to implement single and two qubit gates~\cite{zhangUniversalFastFluxControl2021,barendsCoherentJosephsonQubit2013,reagorDemonstrationUniversalParametric2018}, and to control tunable couplers between qubits~\cite{niskanen_2007_quantumcoherenttunable, bialczak_2011_fasttunablecoupler}. For quantum annealers, flux-based control is essential~\cite{johnson_2010_scalablecontrolsystem} and independent dynamic flux control has been identified as an important resource for quantum enhancement~\cite{khezriCustomizedQuantumAnnealing2021, khezriAnnealpathCorrectionFlux2021,lanting_2017_experimentaldemonstrationperturbative,adameInhomogeneousDrivingQuantum2020,susaQuantumAnnealingSpin2018,ohkuwaReverseAnnealingFully2018,albashDiagonalCatalystsQuantum2020}. 

Previous work on crosstalk calibration has been mainly focused on characterizing and suppressing microwave crosstalk and parasitic qubit couplings~\cite{gambettaCharacterizationAddressabilitySimultaneous2012,feiTunableCouplingScheme2018,zhaoHighContrastInteractionUsing2020,sheldonProcedureSystematicallyTuning2016,pattersonCalibrationCrossResonanceTwoQubit2019,mundadaSuppressionQubitCrosstalk2019,winickSimulatingMitigatingCrosstalk2020}. Although existent since some of the earliest implementations of on-chip flux bias lines~\cite{plourdeFluxQubitsReadout2005,grajcarFourQubitDeviceMixed2006a}, DC flux crosstalk has attracted less attention. In recent gate-based quantum computing implementations, the relative crosstalk, given by the ratio between the coupling from a bias line to an unintended loop and the targeted loop, is on the order of a few percent \cite{abramsMethodsMeasuringMagnetic2019,kounalakisTuneableHoppingNonlinear2018, kellyStatePreservationRepetitive2015,neillBlueprintDemonstratingQuantum2018}. However, for flux-qubit-based quantum annealers, in order to achieve strong inter-qubit interaction, qubits have much larger loop sizes and mutual inductances between them. This leads to much higher relative crosstalk, making crosstalk calibration more challenging.

Commercial quantum annealers rely on magnetic memory elements to feed static flux to qubits and couplers, with crosstalk reduced using suitable integrated superconducting circuit design. Dynamic crosstalk is reduced due to using global control lines~\cite{johnson_2010_scalablecontrolsystem,bunyk_2014_architecturalconsiderationsdesign}. In an alternative quantum annealer implementation using fluxmon qubits \cite{quintanaSuperconductingFluxQubits2017}, the authors outlined a procedure for measuring crosstalk between two coupled fluxmons, based on fitting to analytical circuit models. However, it is unclear whether the method could be easily extended to other systems, where it is hard to obtain an accurate analytical model.

In this work, we introduce an approach to calibrate DC flux crosstalk, which only relies on the symmetries of superconducting circuits. Systematic errors arising from misidentification of the coupling between circuit elements as crosstalk can be reduced by applying this procedure iteratively, with each iteration yielding an improved estimate of the crosstalk matrix elements.
The calibration procedure is automated to allow implementation in large-scale superconducting quantum devices. We used the procedure on two quantum annealing devices, the largest of which has 27 superconducting control loops. We also introduce a method to characterize the error through a different set of measurements, which we executed on the smaller device. The error was measured to be lower than $0.17$\% for all crosstalk coefficients.

The structure of this paper is as follows. We describe our experimental setup in Sec.~\ref{sec:ExperimentSetup}. The method used for crosstalk calibration is presented in Sec.~\ref{sec:CalibrationMethod}. We then discuss the implementation of the method and the automated analysis procedures in Sec.~\ref{sec:CISCIQExperiment}. In Sec.~\ref{sec:ExperimentResults} we present the calibration results, including the measured crosstalk, flux offsets and the error characterization. We compare the measured results with design targets in Sec.~\ref{sec:Comparison}. In Sec.~\ref{sec:Summary} we provide a discussion and conclusions. 

\section{Experimental setup}\label{sec:ExperimentSetup}
\begin{figure*}
\includegraphics[]{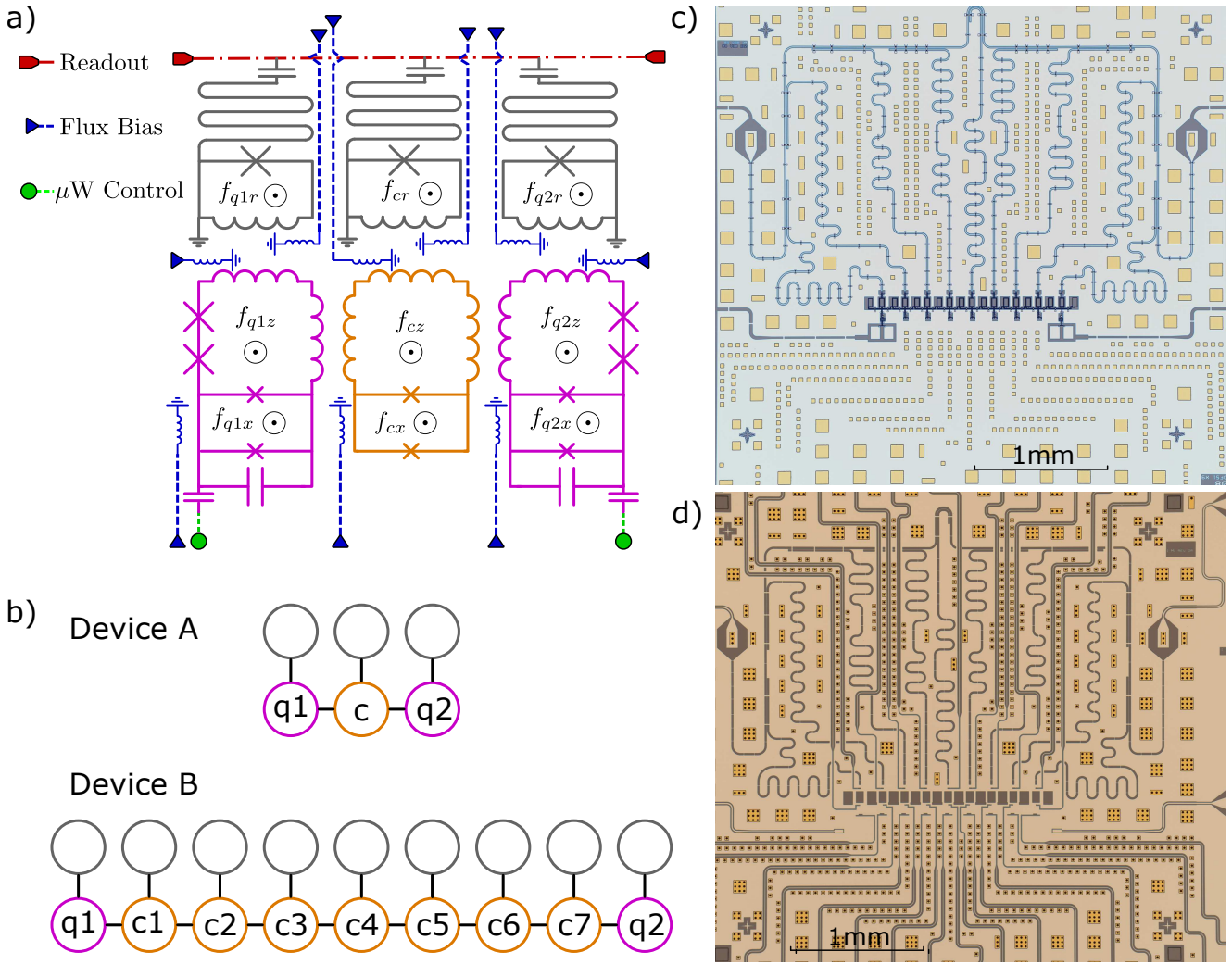}
\caption{\label{fig:DeviceSchematic} (Color online) (a) Circuit schematic of device A. The qubit circuits (left and right, purple) are tunable CSFQs. The coupler circuit (middle, orange) is a tunable rf-SQUID. Each qubit and coupler has two control loops and is coupled to a tunable resonator (top, gray). All resonators are coupled to a joint feedline (top, red). Fluxes in each loop are controlled via the on-chip bias lines (triangles, blue). The design also includes microwave control by capacitively coupling microwave lines (circle, green) to the qubit, allowing for spectroscopy measurement of the device (not used in this work). (b) Diagram representation of devices A and B. Device A contains two qubits (left and right units, purple), the coupler (insider cell, orange) and a tunable resonator for each cell (top, gray). Device B contains two qubits and seven couplers. Device B has the same qubit and coupler circuit schematic and control capabilities as device A, shown in panel (a). (c, d) Microscope image of the qubit and interposer chips of device B. The qubit chip (c) hosts the qubit and readout circuitry and the interposer chip (d) hosts the flux bias lines. The square features (yellow boxes) correspond to indium bumps used to connect the two chips.}
\end{figure*}

We experimentally demonstrate our flux crosstalk protocol on two devices consisting of tunable flux qubits, tunable couplers and flux detectors. These devices are designed to explore high-coherence quantum annealing, based on coupled capacitively-shunted flux qubits (CSFQs)~\cite{weber_2017_coherentcoupledqubits,novikovExploringMoreCoherentQuantum2018a}. 
A circuit schematic of the first device (device A) is shown in Fig.~\ref{fig:DeviceSchematic}(a). It contains two CSFQs and a coupler. Each qubit is formed of a main loop and a secondary loop, named $z$-loop and $x$-loop respectively, in line with their functionality to control the corresponding Pauli terms in the persistent current basis. The coupler has a similar configuration, with a main inductive loop and a secondary split Josephson junction loop. In analogy to the qubit, these loops are named $z$-loop and $x$-loop respectively as well. The $z$-loop of the coupler is inductively coupled to the qubits' $z$-loop, thus acting as a tunable coupler~\cite{harris_2009_compoundjosephsonjunctioncoupler, weber_2017_coherentcoupledqubits}. A set of bias lines is used, with each line designed to couple primarily to a corresponding loop. 

Flux readout devices are coupled to each of the qubits and the coupler. Readout of the persistent current of the qubits is required for standard annealing experiments~\cite{groverFastLifetimePreservingReadout2020, berkley_2010_scalablereadoutsystem}. The additional flux readout of the coupler was added here to aid with calibration of the operation point of the coupler. Each readout device is formed of a tunable rf-SQUID terminating a coplanar waveguide resonator (see Fig.~\ref{fig:DeviceSchematic}(a)), with the rf-SQUID loop coupled to the corresponding  $z$-loop of each qubit or the coupler. The resonators are coupled to a common transmission line. They can be probed by sending a microwave signal at probe frequency $\omega_p$ through the transmission line and measuring the complex transmission coefficient $S_{21}$. In the semi-classical picture, the persistent current in the qubit or coupler $z$-loop generates fluxes threading the resonator rf-SQUID, which changes its effective inductance, leading to a change in resonator's resonance frequency. For a weak enough probing signal, the magnitude of the transmission $|S_{21}|$ has a minimum when $\omega_\text{p}$ coincides with the resonator's resonance frequency.  

Device B consists of two CSFQs coupled by a chain of seven tunable rf-SQUID couplers. Each coupler has its $z$-loop coupled to its neighboring couplers or qubits. The seven couplers act as a coupler chain that mediates flux signals between the end qubits. Fig.~\ref{fig:DeviceSchematic}(b) shows a cartoon representation of this device. It has the potential to realize long-range coupling without trading it off with coherence~\cite{AndrewKerman2021, Daniel2021}.

The devices are fabricated at MIT Lincoln Laboratory, based on the fabrication process outlined in Ref.~\cite{rosenberg3DIntegratedSuperconducting2017}, combining a high coherence chip hosting qubits, an interposer chip, and a multi-layer chip for control and readout wiring. In the work presented here, the devices were realized using only the qubit and interposer chips (see Fig.~\ref{fig:DeviceSchematic} (c,d)), as a preliminary step towards high density annealing circuits including the full three-tier process described in Ref.~\cite{rosenberg3DIntegratedSuperconducting2017, yostSolidstateQubitsIntegrated2020}.

Each device is placed in a sample holder anchored to the mixing chamber plate of a dilution refrigerator. All on-chip flux bias lines are connected to arbitrary waveform generators (AWGs) operating at room temperature through twisted-pair wiring. The connections are appropriately attenuated at room temperature to generate a flux range of a few flux quanta (see Appendix~\ref{app:FridgeWiring} for a complete wiring diagram).

\section{Calibration Method}\label{sec:CalibrationMethod}
\subsection{Defining the crosstalk problem}
The task of crosstalk calibration consists of measuring the coupling from each bias line to each superconducting loop, in order to be able to independently control the external flux applied to each loop. The external flux in each loop normalized to the flux quantum $\Phi_0$ is denoted by $f_i$. The set $\{f_i\}$ forms the flux vector $\mathbf{f}$ of size $N$, where $N$ is the total number of independent loops, including those of the qubits, couplers and resonator SQUIDs. Similarly, we define a current vector $\mathbf{I}$ of the same size $N$. Its $j$-th element $I_j$ corresponds to the current through the bias line $j$. The flux vector $\mathbf{f}$ is related to the current vector $\mathbf{I}$ via

\begin{align}
    \mathbf{f} &= \mathbf{M}\mathbf{I} + \mathbf{f}_0, \label{eqn:FluxCurrentRelation}
\end{align}
where $\mathbf{M}$ is the mutual inductance matrix between bias lines and superconducting loops and $\mathbf{f}_0$ is a vector of flux offsets. The flux offsets are caused by trapped fluxes while cooling down the device to the superconducting phase~\cite{stanCriticalFieldComplete2004}, and by flux noise intrinsic in a superconducting system \cite{lanting_2009_geometricaldependencelowfrequency}. It needs to be emphasized that the effective flux biases induced by inductive coupling between superconducting loops are not included in the flux vector. 

Experimentally the bias line currents are controlled via AWG voltages. The external fluxes are related to the voltages via 
\begin{equation}
    \mathbf{f} = \mathbf{C}\mathbf{V} + \mathbf{f}_0\label{eqn:FluxVoltageRelation},
\end{equation}
where $\mathbf{V}$ is the vector formed of the values $\{V_i\}$ of set voltages and  $\mathbf{C}=\mathbf{M}\mathbf{R}^{-1}$ with $\mathbf{R}$ a diagonal matrix with the $i$-th diagonal element $R_i=V_i/I_i$ being the total resistance of each current carrying wire and its source impedance. The objective of the crosstalk calibration is to measure the coupling coefficients $C_{i,j}$ and the flux offsets $f_{0,i}$. 

The devices used in this work have a design commonality, in that, qubits and couplers are similar circuits, each coupled to a flux tunable resonator. It is useful to group each qubit or coupler with its resonator into a unit cell. This makes it so that each cell has three superconducting loops; $z$, $x$ on the qubit or coupler, and $r$ on the resonator rf-SQUID. We use $C_{p\alpha,q\beta}$ to represent the flux to voltage ratio between the bias line $\beta$ in cell $q$ and the loop $\alpha$ in cell $p$, where $\alpha, \beta \in \{z, x, r\}$ and $p, q \in [1 , m]$. Here $m$ is the total number of unit cells, given by $m=3(9)$ for device A(B). The flux offset in loop $\alpha$ in cell $p$ is denoted by $f_{0,p\alpha}$. Similarly, $f_{p\alpha}$ ($V_{p\alpha}$) represents the flux (voltage) on loop (bias line) $\alpha$ in cell $p$.  This double index notation facilitates the analysis of this circuit, where each readout resonator is nominally coupled to each qubit or coupler. However, the methods discussed below are applicable to more general circuits. Note that we continue to sometimes use the single index notation ($V_i$, $f_i$ and $C_{i,j}$) below. The difference between the single and the double index notations should be clear from the context. 

In anticipation of the iterative method used for calibration, it is useful to distinguish between the estimated values and the true values. We use a prime to denote the estimated value for a specific quantity. For example, $C_{p\alpha,q\beta}^\prime$ refers to the estimated value for $C_{p\alpha,q\beta}$. With the estimated coupling matrix and the estimated flux offsets, we can define an estimated flux vector $\mathbf{f}^\prime$ as 

\begin{align}
    \mathbf{f}^\prime= \mathbf{C}^\prime\mathbf{V} + \mathbf{f}_0^\prime.
\end{align}
Distinguishing between estimated values and the corresponding variables is very important in particular in the discussion of the iterative procedure below. 

\subsection{General approach}
Without assuming any particular model for the superconducting circuit, there are two fundamental symmetry properties that we can exploit for flux crosstalk calibration. Firstly, the device behavior is periodic in one flux quantum with respect to the external flux in each loop. Secondly, the device possesses mirror symmetry with respect to the chip plane. This means the device properties should be the same whether the external fluxes are at $\mathbf{f}$ or $-\mathbf{f}$. This suggests that we can treat crosstalk calibration as an optimization problem. The parameters are the estimated coupling matrix and flux offsets $\mathbf{C}^\prime, \mathbf{f}^\prime_0$, and the target is to maximize some function that measures the periodicity and mirror symmetry of the device behaviour, with respect to the estimated flux coordinates $f_i^\prime$. However, it is not immediately clear whether we can quantify such symmetry properties easily for large-scale devices without taking exceedingly large amounts of data. The convergence of the optimization result could also be very sensitive to the choice of optimization algorithm. For these reasons we leave this to be explored in the future and focus on a more structured and tractable approach in this work. 

The usual approach to flux crosstalk calibration is to identify measurable properties of the circuit that are independent of external fluxes in all but one single loop~\cite{abramsMethodsMeasuringMagnetic2019, kounalakisTuneableHoppingNonlinear2018,neillBlueprintDemonstratingQuantum2018, kellyStatePreservationRepetitive2015}. For example, this could be the resonance frequency of a readout resonator coupled to a tunable qubit. By measuring its response as a function of each control bias voltage $V_j$, while keeping other voltages constant, the matrix element $C_{i,j}$ can be deduced by extracting the periodicity of the measured property as a function of $V_j$. This approach implicitly assumes that the flux in each loop is dominated by the external bias line contributions whereas the fluxes generated by other parts of the circuit are negligible. 

The above assumption does not apply to the devices we are interested in. This is because the interactions between qubits, couplers, and readout resonators are not negligible. The readout resonator frequency depends strongly on the flux bias of its own loop, the qubit or coupler it is coupled to, and to a lesser degree on other units. To address this challenge, we propose an iterative approach to calibration. In the first iteration, it is assumed that when one bias voltage is changed, it only changes the external flux of the loop being addressed, while changes in external fluxes in other loops do not cause any appreciable change in the measured quantity. Using the estimates from the first iteration, subsequent iterations can improve the accuracy of these estimates. In the sections below, we first introduce the procedure for one iteration and then discuss how further iterations are carried out. 

\subsection{CISCIQ}\label{sec:CISCIQ}
We devise a procedure to obtain estimates of the coupling matrix named CISCIQ (an acronym for ``crosstalk into SQUIDs, crosstalk into qubit''). In general terms, it consists of first measuring the coupling elements between bias lines and the SQUID detectors, which is subsequently used to keep the resonator SQUID at nearly fixed operation points, as needed in order to maintain a consistent level of sensitivity to changes in the states of qubits and couplers induced by external biases. This procedure has four stages, discussed below.

\paragraph*{Stage 1.}In the first stage, the resonator direct bias element $C_{pr,pr}$ is measured for each unit $p$. Ignoring the resonator interaction with other quantum elements (qubits, couplers, or other resonators), the resonator frequency is periodic with respect to its own bias line control voltage. By measuring the resonator spectrum as a function of its own bias voltage, we can extract its periodicity, denoted by $P_{pr}$, and the voltage coordinate corresponding to zero flux in the resonator SQUID, denoted by $V_{pr}^*$. Based on these quantities, we estimate
\begin{align}
    C_{pr,pr}^\prime = \frac{1}{P_{pr}},\,\text{and}\ 
    f_{0,pr}^\prime = - \frac{V_{pr}^*}{P_{pr}}.
\end{align}

\paragraph*{Stage 2.}In the second stage, the crosstalk coefficients $C_{pr,q\alpha}$ for all $p, q, \alpha$ are measured. A similar measurement as in Stage 1 is carried out. All voltages except $V_{q\alpha}$ and $V_{pr}$ are set to zero. The readout response is measured as a function of $V_{pr}$ for a set of different values of $V_{q\alpha}$.  The added flux in the probed resonator due to crosstalk from $V_{q\alpha}$ shifts the resonator frequency as a function of $V_{pr}$. For $\delta V_{pr}$ amount of shift per $\delta V_{q\alpha}$, the crosstalk element is given by

\begin{align}
    C_{pr,q\alpha}^\prime = -C_{pr,pr}^\prime \frac{\delta V_{pr}}{\delta V_{q\alpha}} .
\end{align}

\paragraph*{Stage 3.}After Stage 2, we have control of the fluxes in the resonator SQUIDs from all the bias lines. In Stage 3, the $3 \times 3$ sub-matrix formed of the elements $C_{p\alpha,p\beta}$ and the flux offsets $f_{0,pz}, f_{0,px}$ for each unit $p$ are measured. In the remainder of this stage, $V_{q\alpha}$ for all $q\neq p$ and all $\alpha$ are set to zero. To simplify notation, the subscript denoting the cell index is dropped since we are only concerned with intra-unit crosstalk. With this simplified notation we write

\begin{align}
    \begin{pmatrix}
    f_z\\
    f_x\\
    f_r
    \end{pmatrix}
    &=\begin{pmatrix}
    C_{z,z}&C_{z,x}&C_{z,r}\\
    C_{x,z}&C_{x,x}&C_{x,r}\\
    C_{r,z}&C_{r,x}&C_{r,r}
    \end{pmatrix}
    \begin{pmatrix}
    V_z\\
    V_x\\
    V_r
    \end{pmatrix}
    +\begin{pmatrix}
    f_{0,z}\\
    f_{0,x}\\
    f_{0,r}
    \end{pmatrix}.
\end{align}

This stage consists of two measurements. In \textit{measurement (a)}, we measure the fluxes in the qubit or coupler, $f_z$ and $f_x$ by measuring the coupled resonator's transmission at a fixed frequency, while sweeping $x$ and $z$ bias voltages, $V_x, V_z$. During the measurement, the resonator flux bias needs to be fixed, leading to a constraint on the resonator bias voltage,
 
\begin{align}
f_r^\prime = \sum_\alpha C_{r,\alpha}^\prime V_\alpha + f_{0,r}^\prime=0\label{eqn:Stage3ResConstraint},
\end{align}
where $f_r^\prime$ is the approximate resonator flux from external sources, given by the estimates from Stages 1 and 2. Note that the errors in the estimates of Stage 1 and 2 leads to uncompensated crosstalk into the resonator, which can affect the measured transmission, in addition to the changes in transmission due to the changes in $f_z$ and $f_x$. To avoid this complication, we choose to fix $f_r^\prime$ to zero, which makes the resonator first order insensitive to the residual uncompensated crosstalk, allowing us to associate the change in transmission solely with changes in $f_z$ and $f_x$. While setting the resonator flux bias away from zero can increase the overall interaction strength between the resonator and qubit or coupler, potentially leading to more sensitive measurement, we empirically find the benefits of avoiding complication due to uncompensated crosstalk outweighs the cost of slightly weaker sensitivity.

Since $V_r$ is constrained to satisfy the requirements on $f_r^\prime$, the 3-dimensional voltage and flux space is reduced to an effective 2-dimensional relation, such that 
    \begin{align}
        \begin{pmatrix}
        f_z \\
        f_x
        \end{pmatrix}
        =\left(
        \begin{array}{cc}
        C_{z,z}^{\text{eff}} & C_{z,x}^{\text{eff}}\\
        C_{x,z}^{\text{eff}} & C_{x,x}^{\text{eff}}\\
        \end{array}
        \right)
        \begin{pmatrix}
        V_z \\
        V_x
        \end{pmatrix}
        +\begin{pmatrix}
        f_{0,z}^{\text{eff}}\\
        f_{0,x}^{\text{eff}}\\
        \end{pmatrix}.\label{eqn:Effective2DMatrix}
    \end{align}
Specifically, the effective matrix and offsets are related to the actual matrix elements and offsets via
\begin{align}
    C^{\text{eff}}_{z,z} &= C_{z,z}-\frac{C^\prime_{r,z}C_{z,r}}{C^\prime_{r,r}}\label{eqn:EffectiveCzz},\\
    C^{\text{eff}}_{z,x} &= C_{z,x}-\frac{C^\prime_{r,x}C_{z,r}}{C^\prime_{r,r}},\\
    C^{\text{eff}}_{x,z} &= C_{x,z}-\frac{C^\prime_{r,z}C_{x,r}}{C^\prime_{r,r}},\\
    C^{\text{eff}}_{x,x} &= C_{x,x}-\frac{C^\prime_{r,x}C_{x,r}}{C^\prime_{r,r}},\\
    f^{\text{eff}}_{0,z} &= f_{0,z}+\frac{C_{z,r}}{C_{r,r}^\prime}(f_r^\prime-f^\prime_{0,r})\label{eqn:Effectivef0z}~\text{and}\\
    f^{\text{eff}}_{0,x} &= f_{0,x}+\frac{C_{x,r}}{C_{r,r}^\prime}(f_r^\prime-f_{0,r}^\prime).\label{eqn:Effectivef0x}
\end{align}

\textit{Measurement (a)} can be shown to have point reflection symmetry about every half-integer flux point due to both symmetries about the chip plane and periodicity in external fluxes (see Appendix~\ref{app:FluxSymmetry}). These points form a lattice and allow us to find the affine transformation defined by the effective matrix $\mathbf{C}^{\text{eff}}$ and the effective flux offsets $f_{0,z}^{\text{eff}}, f_{0,x}^{\text{eff}}$.  However, as implied by Eqs.~(\ref{eqn:EffectiveCzz})-(\ref{eqn:Effectivef0x}), knowing the effective matrix and offsets is insufficient to determine the complete set of coupling coefficients for this unit. Hence another measurement is needed.

Stage 3 \textit{measurement (b)} repeats \textit{measurement (a)}, but setting $f_r^\prime=\pm 1$. Since the resonator flux is changed by 1 $\Phi_0$, due to the flux periodicity, the resonator response remains the same, up to some translation in the $V_x, V_z$ coordinates. Such translations could be understood in terms of the change in effective flux offsets of the $z(x)$ loop, $f^{\text{eff}}_{0,z(x)}$ due to crosstalk from the resonator bias line. From Eqs.~(\ref{eqn:Effectivef0z}) and (\ref{eqn:Effectivef0x}), we can write the change in effective flux offset $\delta f^{\text{eff}}_{0,z(x)}$ due to change in resonator flux $\delta f_r^\prime$ as,
\begin{align}
    \delta f^{\text{eff}}_{0,z(x)} = \frac{C_{z(x),r}}{C_{r,r}^\prime}\delta f_r^\prime.
    \label{eqn:effectiveFluxOffsetsShifts}
\end{align}
Measurement of the offset shift $\delta f^{\text{eff}}_{0,z(x)}$ is used to determine  $C_{z(x),r}/C_{r,r}^\prime$, which in combination with Eqs.~(\ref{eqn:EffectiveCzz})-(\ref{eqn:Effectivef0x}) enables identifying all the coupling elements and flux offsets in a unit cell.

To extract the offset shifts $\delta f^{\text{eff}}_{0,z}$ and $\delta f^{\text{eff}}_{0,x}$, an effective procedure is to rely on the shifts of the measured two-dimensional datasets quantified along the $V_{z}$ and $V_{x}$ coordinates. These shifts are denoted by $\delta V_{z}$ and $\delta V_{x}$ respectively and are related to the effective offset shifts via

\begin{align}
    \begin{pmatrix}
    \delta f^{\text{eff}}_{0,z}\\
    \delta f^{\text{eff}}_{0,x}
    \end{pmatrix}
    =\begin{pmatrix}
    (C_{z,z}^{\text{eff}})^\prime & (C_{z,x}^{\text{eff}})^\prime\\
    (C_{x,z}^{\text{eff}})^\prime & (C_{x,x}^{\text{eff}})^\prime
    \end{pmatrix}
    \begin{pmatrix}
    -\delta V_{z}\\
    -\delta V_{x}
    \end{pmatrix}.
    \label{eqn:unitVoltageToFlux}
\end{align}
Combining Eqs.~(\ref{eqn:effectiveFluxOffsetsShifts}) and (\ref{eqn:unitVoltageToFlux}) gives
\begin{align}
    \begin{pmatrix}
    C_{z,r}^\prime\\
    C_{x,r}^\prime
    \end{pmatrix}
    =C_{r,r}^\prime
    \begin{pmatrix}
    (C_{z,z}^{\text{eff}})^\prime & (C_{z,x}^{\text{eff}})^\prime\\
    (C_{x,z}^{\text{eff}})^\prime & (C_{x,x}^{\text{eff}})^\prime
    \end{pmatrix}
    \begin{pmatrix}
    -\frac{\delta V_z}{\delta f_r^\prime}\\
    -\frac{\delta V_x}{\delta f_r^\prime}
    \end{pmatrix}.
\end{align}
Then by inverting Eqs.~(\ref{eqn:EffectiveCzz})-(\ref{eqn:Effectivef0x}), the actual 2-dimensional qubit or coupler coupling and offsets can be written in terms of the effective matrix elements and offsets,

\begin{align}
    \begin{pmatrix}
    C_{z,z}^\prime & C_{z,x}^\prime \\
    C_{x,z}^\prime & C_{x,x}^\prime
    \end{pmatrix}
    &=\begin{pmatrix}
    (C_{z,z}^{\text{eff}})^\prime & (C_{z,x}^{\text{eff}})^\prime&\frac{C_{z,r}^\prime}{C_{r,r}^\prime}\\
    (C_{x,z}^{\text{eff}})^\prime & (C_{x,x}^{\text{eff}})^\prime& \frac{C_{x,r}^\prime}{C_{r,r}^\prime}
    \end{pmatrix}
    \begin{pmatrix}
    1 & 0\\
    0 & 1\\
    C^\prime_{r,z} & C^\prime_{r,x}\\
    \end{pmatrix},\\
    f_{0,z}^\prime &= (f_{0,z}^{\text{eff}})^\prime + \frac{C_{z,r}^\prime}{C_{r,r}^\prime}f_{0,r}^\prime~\text{, and}\\
    f_{0,x}^\prime &= (f_{0,x}^{\text{eff}})^\prime + \frac{C_{x,r}^\prime}{C_{r,r}^\prime}f_{0,r}^\prime.
\end{align}

To summarize Stage 3, \textit{measurement (a)} allows us to estimate the effective $2 \times 2$ matrix $\mathbf{C}^{\text{eff}}$ and the effective flux offsets $f^{\text{eff}}_{0,z}, f^{\text{eff}}_{0,x}$, and \textit{measurement (b)} extracts the shifts of the resonator response as a function of $V_z, V_x$. Together with the results from Stages 1 and 2, they complete the flux offsets and $3 \times 3$ block diagonal matrix in each unit cell. 

\paragraph*{Stage 4.} In the last stage, all the remaining crosstalk elements are measured. This is done by performing a measurement similar to Stage 3 \textit{measurement (b)} for each cell $p$, but this time stepping an out-of-cell bias voltage $V_{q\alpha}$.  To fix the resonator flux bias during the measurement, the resonator bias voltage for cell $p$, $V_{pr}$ is constrained so that
\begin{align}
    f_{pr}^\prime = \sum_{\beta\in\{z,x,r\}}C_{pr,p\beta}^\prime V_{p\beta} + C_{pr,q\alpha}^\prime V_{q\alpha} + f_{0,pr}^\prime = 0.
\end{align}
Changes in $V_{q\alpha}$ induce shifts of the measured response in the $V_{pz}, V_{px}$ plane. To find the relation between the crosstalk elements and the shifts, we consider the change in the $z$ and $x$ fluxes in cell $p$, $\delta f_{pz}$ and $\delta f_{px}$ due to changes in the bias voltages $\delta V_{pz}, \delta V_{px}$ and $\delta V_{q\alpha}$. They are given by
\begin{align}
    \begin{pmatrix}
    \delta f_{pz} \\
    \delta f_{px}
    \end{pmatrix}
    =\left(
    \begin{array}{ccc}
    C_{pz,pz}^{\text{eff}} & C_{pz,px}^{\text{eff}} & C_{pz,q\alpha}^{\text{eff}}\\
    C_{px,pz}^{\text{eff}} & C_{px,px}^{\text{eff}} & C_{px,q\alpha}^{\text{eff}}\\
    \end{array}
    \right)
    \begin{pmatrix}
    \delta V_{pz} \\
    \delta V_{px}\\
    \delta V_{q\alpha}
    \end{pmatrix},\label{eqn:InterCrosstalk}
\end{align}
where we introduced $C_{pz,q\alpha}^{\text{eff}}$ and $C_{px,q\alpha}^{\text{eff}}$, given by
\begin{align}
    C_{pz,q\alpha}^{\text{eff}} & = C_{pz,q\alpha} - C_{pz, pr}^\prime\frac{C_{pr,q\alpha}^\prime}{C_{pr,pr}^\prime},~\text{and}\\
    C_{px,q\alpha}^{\text{eff}} &= C_{px, q\alpha} - C_{px, pr}^\prime\frac{C_{pr,q\alpha}^\prime}{C_{pr,pr}^\prime}.\label{eqn:EffInterCrosstalkDef}
\end{align}
A change $\delta V_{q\alpha}$
in the out-of-cell bias voltage $q\alpha$ induces shifts $\delta V_{pz}, \delta V_{px}$ in the resonator response. The effective crosstalk elements can be found by setting $\delta f_{pz}=\delta f_{px}=0$ in Eq.~(\ref{eqn:InterCrosstalk}), yielding
\begin{align}
    \begin{pmatrix}
    \left(C_{pz,q\alpha}^{\text{eff}}\right)^\prime\\
    \left(C_{px,q\alpha}^{\text{eff}}\right)^\prime
    \end{pmatrix}
    &=\left(
        \begin{array}{cc}
        (C_{pz,pz}^{\text{eff}})^\prime & (C_{pz,px}^{\text{eff}})^\prime\\
        (C_{px,pz}^{\text{eff}})^\prime & (C_{px,px}^{\text{eff}})^\prime\\
        \end{array}
        \right)
    \begin{pmatrix}
    -\frac{\delta V_{pz}}{\delta V_{q\alpha}}\\
    -\frac{\delta V_{px}}{\delta V_{q\alpha}}
    \end{pmatrix}.\label{eqn:EffInterCrosstalkValue}
\end{align}
Finally, combining Eqs.~(\ref{eqn:EffInterCrosstalkDef}) and (\ref{eqn:EffInterCrosstalkValue}) gives
\begin{align}
    \begin{pmatrix}
    C_{pz,q\alpha}^\prime\\
    C_{px,q\alpha}^\prime
    \end{pmatrix}
    &=\left(
        \begin{array}{cc}
        (C_{pz,pz}^{\text{eff}})^\prime & (C_{pz,px}^{\text{eff}})^\prime\\
        (C_{px,pz}^{\text{eff}})^\prime & (C_{px,px}^{\text{eff}})^\prime\\
        \end{array}
        \right)
    \begin{pmatrix}
    -\frac{\delta V_{pz}}{\delta V_{q\alpha}}\\
    -\frac{\delta V_{px}}{\delta V_{q\alpha}}
    \end{pmatrix}\nonumber\\
    &+\begin{pmatrix}
    C_{pz, pr}^\prime\frac{C_{pr,q\alpha}^\prime}{C_{pr,pr}^\prime}\\
    C_{px, pr}^\prime\frac{C_{pr,q\alpha}^\prime}{C_{pr,pr}^\prime}
    \end{pmatrix}.\label{eqn:Stage4Crosstalk}
\end{align}

\subsection{Limitations of CISCIQ}
The CISCIQ procedure relies on assuming that when a circuit element (a qubit, coupler, or SQUID) is measured, its properties as a function of an externally applied flux are negligibly affected by the interaction with other circuit elements. For example, in Stage 1, it is assumed that the tunable resonator frequency is periodic in its own bias voltage. However, due to crosstalk from the resonator bias line to the coupler loop, the coupler properties change over the range of the resonator bias swept, which in turn changes the resonator due to their inductive interaction. In Stage 2, while the coupler $z$ bias is swept, its resonator SQUID flux changes not only because of the finite crosstalk from the $z$ bias line to the SQUID loop but also because of the state of the coupler changing. 

\begin{figure}
    \centering
    \includegraphics[]{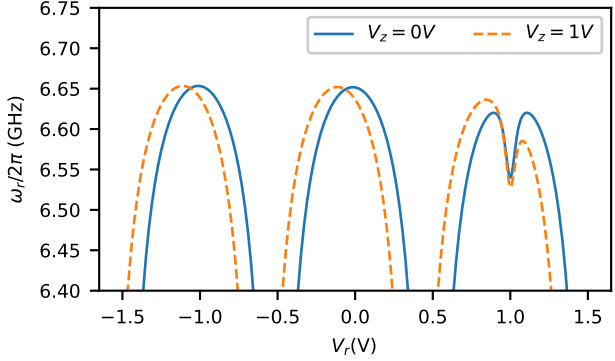}
    \caption{(Color online) The simulated readout resonator spectrum as a function of its bias voltage, using the inductive loading model to describe the interaction between the resonator and the coupler (see text). The solid (blue) and orange (dashed) lines correspond to the coupler coupled to the resonator being biased at two different $V_z$ values, of $0$ and $1$ Volt respectively. The coupling coefficients and flux offsets used for this model are $C_{z,z}=C_{r,r}=1~\Phi_0/V, C_{r,z}=C_{z,r}=0.1~\Phi_0/V, f_{0,r}=0, f_{0,z}=0.4$, which are realistic in our devices. Translational symmetry is broken due to the coupler resonator interaction.
    \label{fig:NonLinearCrosstalkSemiClassical}}
\end{figure}

To illustrate the expected role of circuit interactions, we use a simple model to calculate the effect of the resonator-coupler interaction on the resonator response. The interaction is modeled via the inductive loading of the rf-SQUID inductance by the coupler effective quantum inductance. We consider a single coupler-resonator cell, and assume only coupling between the $r$ and $z$ bias lines and the resonator and coupler loops. By numerically finding the coupler circuit effective inductance and solving the classical rf-SQUID equation, the resonator spectrum in terms of the resonator and coupler $z$ bias can be calculated (see Appendix~\ref{app:CouplerResonator} for additional details). Figure~\ref{fig:NonLinearCrosstalkSemiClassical} shows the resonator spectrum as a function of its bias voltage $V_r$, for two different values of coupler $z$ bias voltage $V_z$. The dominant feature is the resonator frequency change due to change in its own bias. However, due to crosstalk, the coupler flux bias also changes as a function of $V_r$, which changes its inductive loading effect on the resonator and hence the resonator frequency. Therefore the translational symmetry in $V_r$ that is used for calibration in Stages 1 and 2 is broken. We note here that the inductive loading model does not capture the full interaction between the resonator and coupler, rather it serves as an example to highlight the increased complexity of calibration due to strong interactions between circuit elements. 

The above analysis can be extended to other stages of CISCIQ, where the coupler effective inductance loads the qubit and the coupler ground state current acts as an effective bias seen by the qubit. Based on the single qubit and coupler persistent current and their mutual inductances, such an effect could produce an error of tens of  $\text{m}\Phi_0$ (see Appendix~\ref{app:CircuitParameters}). While it is possible to develop models to capture the interaction for small systems, developing an accurate model for a large system is a daunting task. Apart from the qubit-coupler, qubit-resonator or coupler-resonator interactions, since the different resonators are coupled to a single transmission line for readout, resonance collisions also distort the readout signal. Larger devices are particularly prone to this problem. Therefore, CISCIQ alone does not provide a good enough measurement of the coupling matrix and flux offsets.

\subsection{CISCIQi}
To reduce the errors in calibration coefficients found with CISCIQ, which are due to the systematic effects discussed in the previous subsection, we propose an iterative approach, abbreviated as CISCIQi. With this procedure, the measured coupling coefficients and flux offsets from CISCIQ are taken as an initial estimate. Further iterations of CISCIQ are carried out to gradually improve the initial estimates. In each iteration, one controls directly not the voltages but a set of new coordinates - which are the fluxes applied to the loops, calculated based on the estimated coupling coefficients and offsets. More specifically, the first iteration can be considered as giving the estimates of $\mathbf{C}$ and $\mathbf{f}_0$ as $\mathbf{C}^{(1)\prime}$ and $\mathbf{f}^{(1)\prime}_0$ respectively. In the absence of interactions between individual qubit, coupler, and resonator circuits, these estimates are accurate, limited only by experimental noise. However this is not the case for the reasons explained in the previous section. Nevertheless, the quantities
\begin{align}
\mathbf{f}^{(1)} = \mathbf{C}^{(1)\prime} \mathbf{V} + \mathbf{f}_0^{(1)\prime},
\end{align} 
where $\mathbf{C}^{(1)\prime}$ and $\mathbf{f}_0^{(1)\prime}$ are the estimates of the coupling matrix and flux offsets obtained from iteration 1, are a good approximation for fluxes in the loops. Then the flux relation in Eq.~(\ref{eqn:FluxVoltageRelation}) can be recast into the form
\begin{align}
    \mathbf{f}=\mathbf{C}^{(2)}\mathbf{f}^{(1)} + \mathbf{f}_0^{(2)}
\end{align}
where we introduced 
\begin{align}
    \mathbf{C}^{(2)}&=\mathbf{C}\left(\mathbf{C}^{(1) \prime}\right)^{-1}
\end{align}
and 
\begin{align}
    \mathbf{f}_{0}^{(2)}&=-\mathbf{C}\left(\mathbf{C}^{(1) \prime}\right)^{-1} \mathbf{f}_{0}^{(1) \prime}+\mathbf{f}_{0}.
\end{align}

The task in iteration 2 is to estimate $\mathbf{C}^{(2)}$ and $\mathbf{f}_{0}^{(2)}$ by sweeping the components of $\mathbf{f}^{(1)}$ and measuring the circuit response. Because the basis vectors in the estimated flux coordinates $\mathbf{f}^{(1)}$, as compared to those in the voltage coordinates $V$, are closer to the corresponding basis vectors in the real flux coordinates $\mathbf{f}$, the assumptions made in CISCIQ on periodicity with respect to controls are better justified. For example, when repeating Stage 1 during iteration 2, the resonator frequency as a function of $f^{(1)}_{pr}$ has smaller departures from periodicity than $f^{(1)}_{pr}$  as a function of $V_{pr}$ in iteration 1. In Stage 2, one measures the crosstalk from a source bias to a target resonator by stepping the source biases with integers of flux quanta according to the estimated coupling coefficients. By doing this the resonator spectrum better obeys translational symmetry over different source bias settings, because the effect of the interactions with the rest of the circuit is reduced when changes in applied fluxes are close to the circuit periodicity. In Stage 3, the data is expected to have better point reflection symmetry in the $f^{(1)}_{pz}, f^{(1)}_{px}$ plane as compared to $V_{pz}, V_{px}$. In Stage 4, similarly to Stage 2, we choose integer flux quanta steps in the crosstalk source bias to null out modulations within a period. In all stages, resonance collisions become less likely in the second iteration, as the resonators bias points are better controlled, due to the reduced changes in flux coupled from other elements. 

At the $n$-th iteration, this procedure yields the estimates for the coupling matrix $\mathbf{C}^{(n)\prime}$ and the offset vector $\mathbf{f}_0^{(n)\prime}$. Combining the coupling matrix and offsets measured at each iteration, the estimates after $n$ iterations for the coupling matrix and the flux offset are
\begin{align}
\mathbf{C}^\prime &= \mathbf{C}^{(n)\prime}\mathbf{C}^{(n-1)\prime}\dots\mathbf{C}^{(1)\prime}\label{eqn:IterationMatrix}
\end{align}
and 
\begin{widetext}
\begin{align}
\mathbf{f}_0^\prime  &= \mathbf{C}^{(n)\prime}\left(\mathbf{C}^{(n-1)\prime}\left(\dots \left(\mathbf{C}^{(2)\prime}\mathbf{f}_0^{(1)\prime}+\mathbf{f}_0^{(2)\prime}\right)+\dots \right)+\mathbf{f}_0^{(n-1)\prime}\right)+\mathbf{f}_0^{(n)\prime}\label{eqn:IterationOffset}.
\end{align}
\end{widetext}

\subsection{Fast offsets calibration}\label{sec:OffsetCalibration}

The coupling matrix is expected to remain constant in the course of an experiment while the device is kept cold inside a dilution refrigerator. However, the flux offsets change over time due to flux noise and trapped flux. Therefore a time efficient method is desired to recalibrate flux offsets. Here we introduce such a procedure, which relies on the knowledge of the estimate of the coupling matrix, assumed to remain constant. The method makes use of similar measurements as in Stages 1 and 3 of CISCIQ. To measure the offset of the resonator, the corresponding approximate flux coordinate $f_{pr}^\prime$ is swept around 0, and at each flux setting the resonator transmission is measured over a frequency range around resonance. The signal is expected to be mirror reflection symmetric about a value, denoted by  $f_{pr}^{\prime *}$, which corresponds to the flux in the resonator being equal to zero. Based on this, the new estimated flux offset is related to the old estimated flux offset $f_{0,pr}^{\prime}$ by

\begin{align}
    f_{0,pr}^{\prime} \xrightarrow[]{} f_{0,pr}^{\prime} - f_{pr}^{\prime *}\label{eqn:OffsetUpdate}.
\end{align}

Similarly, to find the new offset of the qubit or coupler, the estimated flux coordinates $f_{pz}^\prime, f_{px}^\prime$ are swept while probing resonator transmission close to the peak frequency of the resonator for unit $p$. The signal is point reflection symmetric about some point $(f_{pz}^{\prime *}, f_{px}^{\prime *})$. The new estimated flux offset and the old ones are related analogously to the resonator offset in Eq.~(\ref{eqn:OffsetUpdate}). If this set of measurements reveals that the offset drifts are large, the procedure can be iterated to eliminate the apparent offset shifts due to circuit interactions.

\subsection{The error of the calibration procedure}\label{sec:ErrorMethod}
There are several sources of errors for the CISCIQi calibration procedure. Firstly, all the data collected has noise contributions from the microwave amplifiers, and flux drifts occur while taking the data. Secondly, the fitting algorithms applied to identify translational and point reflection symmetries have estimate errors arising from the finite range and sampling for the collected data. Last and most significantly, systematic errors arise from circuit interactions, which are only partially mitigated even after application of multiple iterations in the calibration procedure. To characterize the errors in the crosstalk matrix and flux offsets considering all the sources of errors is therefore a complex task. Conventional error propagation analysis is not suitable for our method because the output of the analysis, i.e. the periodicities and translations extracted, depends linearly on the coupling coefficients, as well as nonlinearly on the interactions between circuit elements. As an example, Stage 3(b) measurement of CISCIQ relies on couplings measured in Stage 1 and 2. The errors in the resonator bias from Stage 1 and 2 would result in resonator frequencies being different for $f_r^\prime=0, \pm1$ when doing Stage 3 measurements. This causes transmissions at different resonator flux biases not being simply translated versions of each other. The errors in the extracted translations and crosstalk in Stage 3 thus have a substantially nonlinear dependence on the errors from Stage 1 and 2. Given the above consideration, we propose an error characterization method that is motivated by the purpose of crosstalk calibration, which is to gain independent control on each flux bias. 

The error characterization relies on a set of measurements performed to determine to what extent the fluxes can be controlled independently. Ideally, when a change $\Delta f_j^\prime$ in the estimated flux $f_j^\prime$ is applied to loop $j$, the flux in other loops should remain unaffected. Any change in flux can be conveniently measured using the abbreviated offset measurement procedure discussed in Sec.~\ref{sec:OffsetCalibration}. For $i,j\in [1,N], i\neq j$, the quantity 
\begin{align}
     \Theta_{i,j} = \frac{\Delta f_{0,i}^\prime}{\Delta f_j^\prime}
\end{align}
is a measure of the remaining control crosstalk. To measure this quantity in a way that is robust against systematic errors from circuit interactions, one can set $\Delta f_j^\prime$ to be integer flux quanta. This leads to a reduced effect of circuit interaction, due to their periodic dependence on applied fluxes.

\section{Implementation of the CISCIQ method}\label{sec:CISCIQExperiment}
In this section we discus the experimental implementation of the first iteration in the CISCIQi iterative procedure. We discuss this iteration in detail, given that the analysis tools carry over to subsequent iterations. We present measurement for calibration of device A as examples. Measurements on device B are carried out similarly.

\subsection{Stage 1}
In Stage 1, for each resonator the transmission is measured versus the probe frequency and resonator bias voltage. The voltage bias sweep is chosen to cover a few flux quanta, in order to allow determination of the periodicity. We note that in the first iteration we choose bias voltage ranges that are relatively large, to allow determining the period in the presence of relatively strong, uncompensated spurious flux generated by other circuit elements. 

\begin{figure}
    \centering
    \includegraphics[]{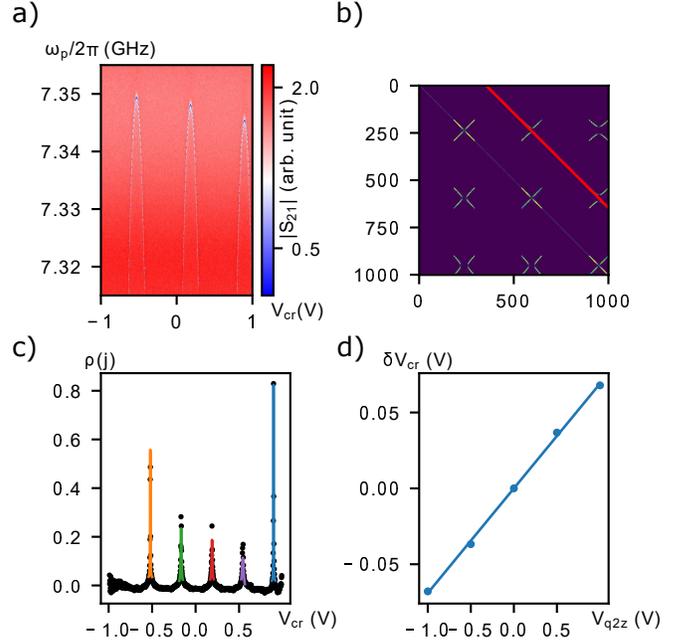}
    \caption{(Color online) (a) Transmission magnitude $|S_{21}|$ versus bias voltage $V_{cr}$ and probe frequency $\omega_{p}$ for the unit cell $c$ of device A. (b) Recurrence plot of image in (a) with detected line (red). The horizontal and vertical axes correspond to pixel indices in the bias voltage dimension in (a). (c) Reflection symmetry correlation coefficient $\rho(j)$ versus bias voltages $V_{cr}$. Each local maximum is locally fitted to a Lorentzian line-shape (orange, green, red, purple and blue curves) to obtain sub-pixel accuracy in the offset. (d) Shifts of resonator response $\delta V_{cr}$ versus crosstalk source bias voltages $V_{q2z}$. The dots are data points and the line is the fit result. \label{fig:RecurrenceAnalysis}}
\end{figure}

Figure~\ref{fig:RecurrenceAnalysis}(a) shows a color plot of the transmission magnitude versus bias voltage and probe frequency. At resonance, the magnitude of transmission has a dip relative to the background. To extract the periodicity and offset, one could extract the resonance frequencies as a function of bias voltage and fit it to the resonator model or a simpler periodic function. However, this method becomes difficult to automate due to the presence of other features in the transmission arising from other readout resonators and spurious package resonances. In addition, fitting the transmission  requires small frequency steps and an analytical transmission model can be hard to obtain when the tunable resonator is driven at high power. Hence, an image processing based method is used instead. The transmission data can be considered as an image with the first dimension being the bias voltage $V_{cr}$, the second dimension being the probe frequency $\omega_p$, and the third dimension being the magnitude of the transmission $|S_{21}|$.  Before extracting the period of the data, edge detection techniques are applied to enhance the resonance features (see Appendix~\ref{app:RecurrenceAnalysis}). To extract the period of the resonator bias, recurrence plot analysis is used \cite{marwanRecurrencePlotsAnalysis2007}. Recurrence plots are a method to visualize symmetries in time series data and are adapted here to identify periodicities within an image and translations between two different images. Given two images $A(i_1,i_2)$ and $B(j_1,j_2)$, the recurrence plot is a new dataset $R(i_1,j_1)$ defined as  
\begin{align}
R(i_1,j_1) = \begin{cases}
1, \text{if} \sqrt{\sum\limits_{i_2,j_2}(A_{i_1,i_2}-B_{j_1,j_2})^2\delta_{i_2,j_2}} \leq \epsilon \\
0, \text{otherwise} \\
\end{cases},
\end{align}
where $\delta_{i_2,j_2}$ is the Kronecker delta and $\epsilon$ is a threshold chosen to maximize the contrast in the recurrence plot (see Appendix~\ref{app:RecurrenceAnalysis}). The arguments in $A$, $B$, and $R$ are integer valued indices. We apply the recurrence plot to the case where both the $A$ and $B$ images are the acquired dataset. The y-intercept of 45-deg lines in the plot corresponds to the amount of translation needed on one of the images to overlap onto the other. We use line detection via the Hough transform to extract the translations from the recurrence plot. An example resonator transmission image and its corresponding recurrence plot are shown in Fig.~\ref{fig:RecurrenceAnalysis}(a) and (b).

To extract the flux offset, the reflection symmetry of the data is analyzed. The measured transmission magnitude $|S_{21}|$ can be considered as an image $A(i_1, i_2)$ of dimensions $(m_1, m_2)$. The correlation coefficients between image $A$ and its reflection about all bias indices are calculated. The reflected image $B(i_1, i_2; j)$ about a particular bias index $j$ is given by
\begin{align}
    B(i_1,i_2;j)=A(2j-i_1,i_2).
\end{align}
The correlation coefficient $\rho(j)$ is given by 
\begin{align}
    &\rho(j) \nonumber\\
    = &\frac{\sum\limits_{i_1, i_2}[A(i_1,i_2)-\bar{A}(j)][B(i_1,i_2;j)-\bar{B}(j)]}{\sqrt{\sum\limits_{i_1, i_2}[A(i_1,i_2)-\bar{A}(j)]^2\sum\limits_{i_1, i_2}[B(i_1,i_2;j)-\bar{B}(j)]^2}},
\end{align}
where the summations range over $i_1\in [\text{Max}(1, 2j-m_1), \text{Min}(m_1,2j-1)]$ and $i_2\in [1,m_2]$, and $\bar{A}(j), \bar{B}(j)$ are the average values of $A,B$ over the same range. The ranges serve to pick out the overlap region between the original and reflected images. The correlation coefficient used here is adapted from the Pearson correlation coefficient applied to samples. It is normalized to lie between $[-1, 1]$, so images with different overlap sizes can be fairly compared. The peaks in $\rho(j)$ correspond to points of reflection symmetry in the image, identified with half and whole integer flux quanta in the resonator loop. Figure \ref{fig:RecurrenceAnalysis}(c) shows the result of this calculation. Finally, the integer flux quanta points can be distinguished from the half integer flux quanta points by checking whether there is a dip in transmission within the frequency range swept, at that bias point.

\subsection{Stage 2}
In Stage 2, for each resonator the transmission is measured versus the resonator direct bias and probe frequency, at a few voltages of each indirect bias line, with all other bias voltages set to zero. Recurrence plots are used to extract the translations in the two dimensional data for each value of the applied indirect bias voltage. The translation versus indirect voltage is fit by a line, whose slope represents the amount of crosstalk.  Figure~\ref{fig:RecurrenceAnalysis}(d) shows an example of such a fit. While the translations versus crosstalk source voltage follow a linear dependence to a good approximation, small systematic errors are observed due to interactions of the resonator with the rest of the circuit.

\subsection{Stage 3}
\begin{figure}
    \centering
    \includegraphics[]{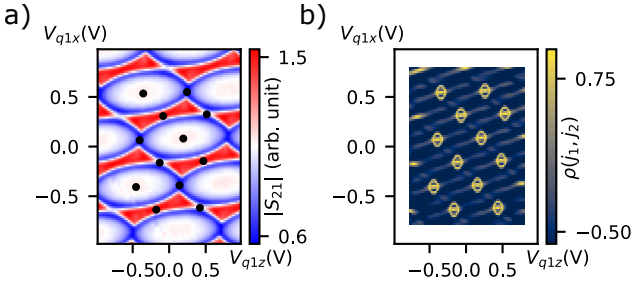}
     \caption{(Color Online) (a) Resonator transmission magnitude $|S_{21}|$ versus $z, x$ bias voltages for unit cell $q1$. The detected inversion symmetry centers are shown by the black dots.  (b) Correlation coefficient $\rho (j_1,j_2)$ for the image in (a). Each local maximum in the image corresponds to one KAZE feature (highlighted circles). We note that for a range up to about half the expected periodicity from the edge of the transmission measurement, the correlation coefficients are not calculated and plotted as white. This is because the correlation becomes an unreliable measure of symmetry near the edge. \label{fig:InversionSymmetry}}
\end{figure}

In Stage 3(a), for each unit cell the resonator transmission is probed at a fixed probe frequency, with the resonator external flux held constant while sweeping its directly coupled qubit or coupler $x$ and $z$ biases. Typically, the probe frequency is below the peak frequency by about half of its linewidth to maximize contrast. The bias ranges are typically swept over two to four periods in both directions and the step size is of the order of $1\,\%$ of the observed periodicity. The resonator flux is kept at zero during the sweep. At this bias point, the resonator is flux-insensitive to first order. This choice of the resonator bias  minimizes the frequency change of the resonator due to residual crosstalk when the $z$ and $x$ biases are swept, thus preventing deterioration of the measurement signal. This measurement generates an image that has point reflection symmetry about integer and half-integer flux points (see Appendix~\ref{app:FluxSymmetry}). The measured transmission magnitude $|S_{21}|$ can be considered as an image $A(i_1,i_2)$ of dimensions $(m_1, m_2)$. To extract the point reflection symmetry centers, the correlation coefficient between the image and the image inverted about some point $(j_1,j_2)$ is calculated. The inverted image $B(i_1,i_2;j_1,j_2)$ is given by 

\begin{align}
    B(i_1,i_2;j_1,j_2)=A(2j_1-i_1,2j_2-i_2).
\end{align}
The correlation coefficient $\rho(j_1,j_2)$ is given by
\begin{widetext}
\begin{align}
\rho{(j_1,j_2)} =\frac{\sum\limits_{i_1, i_2}[A(i_1,i_2)-\bar{A}(j_1,j_2)][B(i_1,i_2;j_1,j_2)-\bar{B}(j_1,j_2)]}{\sqrt{\sum\limits_{i_1, i_2}[A(i_1,i_2)-\bar{A}(j_1,j_2)]^2\sum\limits_{i_1, i_2}[B(i_1,i_2;j_1,j_2)-\bar{B}(j_1,j_2)]^2}},\label{eqn:CorrelationCoeff}
\end{align}
\end{widetext}
where the summations range over $i_{1(2)}\in [\text{Max}(1,2j_{1(2)}-m_{1(2)}),\text{Min}(m_{1(2)},2j_{1(2)}-1)]$, and $\bar{A}(j_1,j_2)$, $\bar{B}(j_1,j_2)$ are the average values of $A$, $B$ over the same range. Local maxima in the image $\rho(i_1, i_2)$ correspond to points with maximum point reflection symmetry. Instead of simple peak detection, KAZE feature recognition~\cite{Alcantarilla12eccv} is applied  to the image $\rho$, which detects blobs in the image. Then, by filtering out features that are not close to any local maximum, the coordinates of the remaining features can then be identified with point reflection symmetry centers. It is empirically found that the KAZE feature detection outperforms simple local maximum detection, in cases where resonator collision causes the measured transmission to deviate from the expected symmetry. The feature detection also allows sub-pixel precision, which removes the need to take time-consuming, high-density measurements. Fig.~\ref{fig:InversionSymmetry}(a) shows the point reflection symmetry centers plotted on the measured data, ordered by their distances to the origin and  Fig.~\ref{fig:InversionSymmetry}(b) shows the point reflection correlation coefficients calculated, with the KAZE features overlaid.

The inversion symmetry centers are coordinates in $z$ and $x$ bias voltages corresponding to half-integer flux quantum in $z$ and $x$ loops. The next task is to identify an affine transformation that converts these inversion symmetry centers to coordinates in external fluxes. In principle, any three inversion symmetry centers that are not co-linear are sufficient to define such a transformation. However, due to various noise sources, it is likely that different choices of inversion centers will lead to slightly different transformations. To make use of the full lattice of inversion symmetry centers, the affine transformation parameters can be treated as fitting parameters. The optimal transformation is found by minimizing the distance between transformed lattice coordinates and the ideal lattice coordinates~\cite{novikovExploringMoreCoherentQuantum2018a}. 

\begin{figure}
    \centering
    \includegraphics{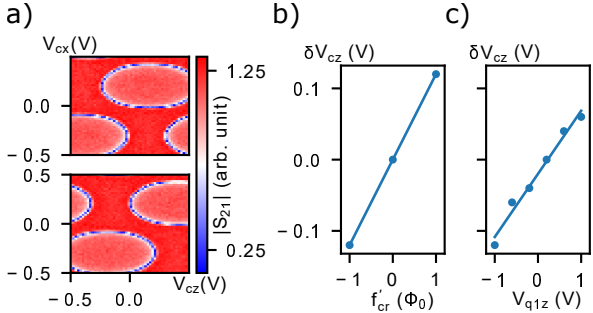}
    \caption{(Color online) (a) Coupler resonator transmission measurement versus $V_{cz}, V_{cx}$, at $f^\prime_{cr}=-1$ (top) and  $f^\prime_{cr}=1$ (bottom). (b) Shifts of the coupler 2d scan in $z$ direction, $\delta V_{cz}$ versus resonator bias $f_{cr}^\prime$. (c) Shifts of the coupler 2d scan in $z$ direction versus crosstalk source $V_{q1z}$. In both (b) and (c) the dots are data points and the line is the fit result.}
    \label{fig:Stage4Analysis}
\end{figure}

In Stage 3(b), for each unit the resonator transmission is probed at a fixed frequency, while sweeping the unit $x$ and $z$ bias voltages and maintaining $f_r^\prime=\pm 1$. Fig.~\ref{fig:Stage4Analysis}(a) shows the measured data for the coupler unit. Translations between images in both the $z$ and $x$ directions are simultaneously extracted using \textit{scikit-image} image registration routine ~\cite{waltScikitimageImageProcessing2014, guizar-sicairosEfficientSubpixelImage2008}. The translations versus resonator bias values are fitted to a line and the slope can be related to the crosstalk value as discussed in Sec.\ref{sec:CISCIQ} Stage 3.

\subsection{Stage 4}
In Stage 4, measurements similar to those in Stage 3(b) are performed. For each unit, the resonator transmission is probed while fixing the resonator flux bias and sweeping the unit's $x$, $z$ biases, and stepping another crosstalk source bias voltage. The ranges for crosstalk source voltage are chosen to cover more than one flux quantum flux bias in the corresponding loop, to ensure the translations measured are not biased due to the modulation of circuit interaction within a period. An example of translations extracted versus crosstalk source bias voltages and the corresponding fits are shown in Fig.~\ref{fig:Stage4Analysis}(c). The slope of the line fit is then related to the crosstalk coupling via Eq.~(\ref{eqn:Stage4Crosstalk}).

\section{Calibration results for devices A and B}\label{sec:ExperimentResults}
\subsection{CISCIQ\lowercase{i}}
After the application of CISCIQ as discussed above, further iterations are performed by sweeping the approximate flux coordinates, where fluxes are calculated according to the estimates of the coupling matrix and flux offset obtained in the previous iteration. Three complete iterations are completed for both devices A and B. The estimated values of coupling coefficients at the end of each iteration are calculated using Eq.~(\ref{eqn:IterationMatrix}).

\begin{figure*}
    \centering
    \includegraphics{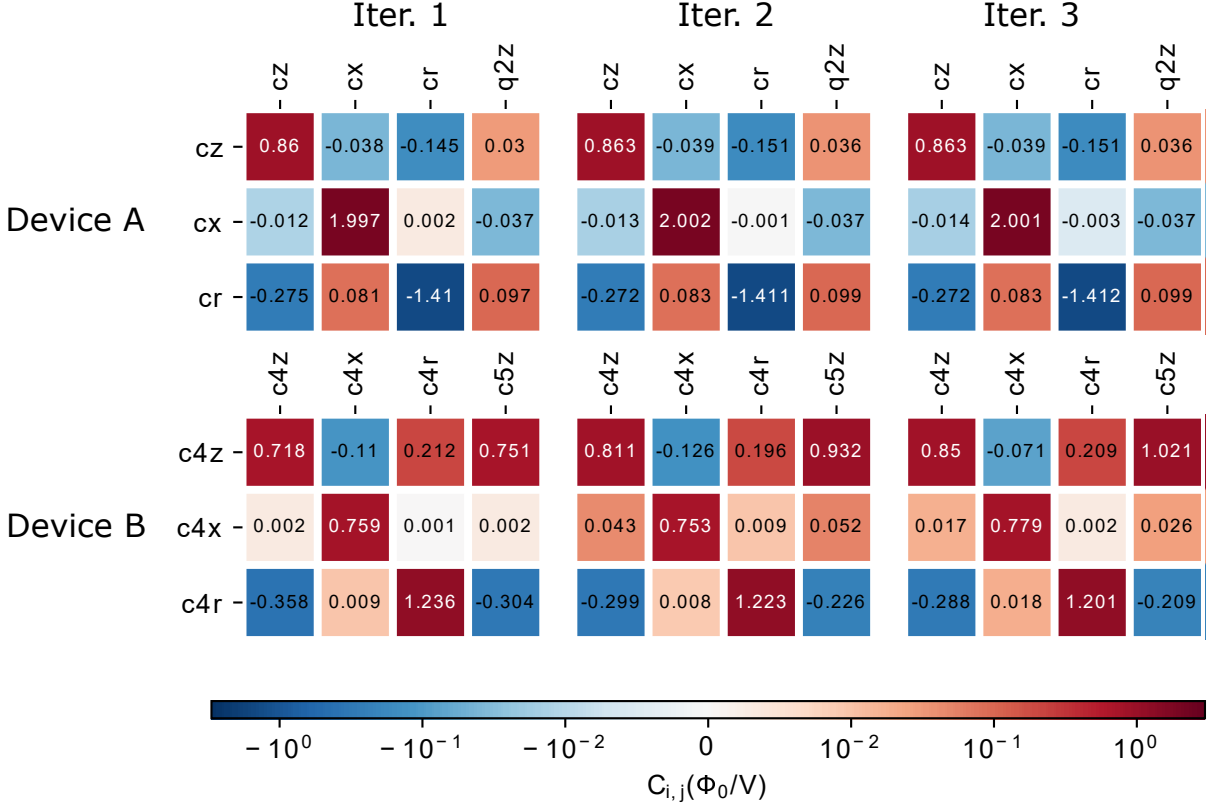}
    \caption{(Color online) A selection of the coupling coefficients for each iteration for devices A (top) and B (bottom). From left to right they correspond to iteration 1, 2 and 3. For each image the column and row labels correspond to source and target loops, respectively.}
    \label{fig:IterationSubmatrices}
\end{figure*}

\begin{figure}
    \centering
    \includegraphics{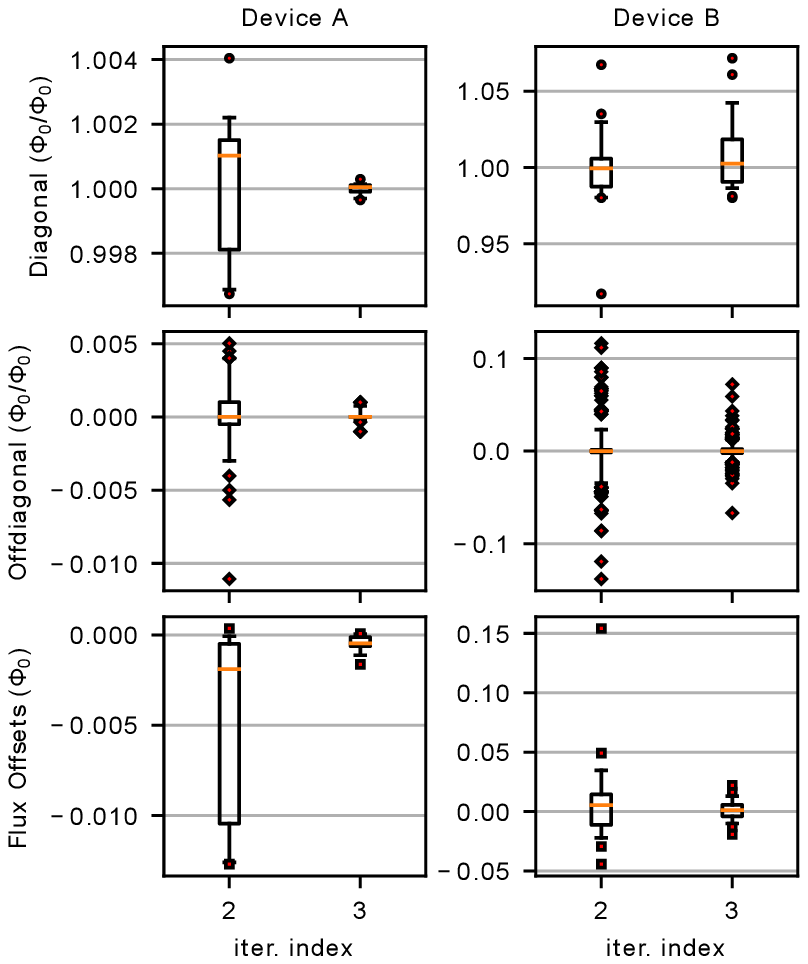}
    \caption{(Color online) For device A (left), B (right) respectively, the statistical box plots of the diagonal (top), off-diagonal (middle) coupling coefficients in $\mathbf{C}^{(n)\prime}$, and the flux offsets (bottom) in  $\mathbf{f}^{(n)\prime}_0$, for iteration 2 and 3. The orange bar is the median, the black box correspond to the lower and upper quartiles, the segments contain 5 to 95 percentile of the data and the dots are outliers. For device B, the coupling coefficients and offsets corresponding to the $c2$ unit are excluded in the plots due to device failure (see text).}
    \label{fig:ConvergencePlot}
\end{figure}

To illustrate how the coupling elements change with iteration, we show in Fig.~\ref{fig:IterationSubmatrices} a subset of the coupling matrix elements for iterations 1-3 for devices A and B. Note that the coupling elements shown correspond to units at the center of devices A and B, which are most affected by systematic crosstalk errors due to interactions with other circuit elements. We observe that all the coupling elements change, with typically a smaller change between iteration 2 and 3 than between iteration 1 and 2. To further illustrate the effectiveness of the iterations, in Fig.~\ref{fig:ConvergencePlot}(a) we show the statistical box plots of coupling coefficients and flux offsets in $\mathbf{M}^{(n)\prime}, \mathbf{f}^{(n)\prime}$ for $n=2, 3$. It is clear from the plot that the coupling matrices in iterations 2 and 3 are approaching identity, and the flux offsets are approaching zero. 

For device A, it is worth noting that the corrections in off-diagonal matrix elements in iteration 2 are about $10\,\text{m}\Phi_0 / \Phi_0$, and the correction in flux offsets are about $10\,\text{m}\Phi_0$. Assuming $60\,\text{pH}$ mutual inductance between circuit elements, which is the typical value, the required persistent current to generate $10\,\text{m}\Phi_0$ flux is $0.34\,\mu\text{A}$. This number is comparable to the maximum ground state current in the coupler $z$-loop, which is $0.45\,\mu\text{A}$. As the persistent current gets modulated by flux bias, corrections on the order of $10\,\text{m}\Phi_0 / \Phi_0$ in iteration 2 are consistent with the level of interactions between circuit elements (see Appendix \ref{app:CircuitParameters} for a more detailed comparison). In iteration 3, the matrix element corrections are below $2\,\text{m}\Phi_0/\Phi_0$ and the flux offsets corrections are below $2\,\text{m}\Phi_0$. Since the flux drifts measured (see Sec.~\ref{sec:OffsetDriftsData}) are also about $2\,\text{m}\Phi_0$, this suggests that further iterations would be limited by random flux jumps and not improve the calibration measurement much further. 

When compared to device A, device B has three times as many control loops. In addition, device B has overall stronger circuit interactions,  because the couplers are designed to have about three times as large a persistent current and flux sensitivity compared to the qubit (see Appendix~\ref{app:CircuitParameters}). This is compounded by the fact that more resonators are on the same feedline in device B, leading to increased errors in resonator readout. Besides, device B also suffered from a partial device failure: the resonator SQUID in cell $c2$ could not be tuned. This cell, including the resonator and the coupler, remained uncalibrated during the CISCIQi procedure (additional techniques were used to calibrate this unit for other experiments, which we do not discuss in this work). 

\begin{figure}
    \centering
    \includegraphics[]{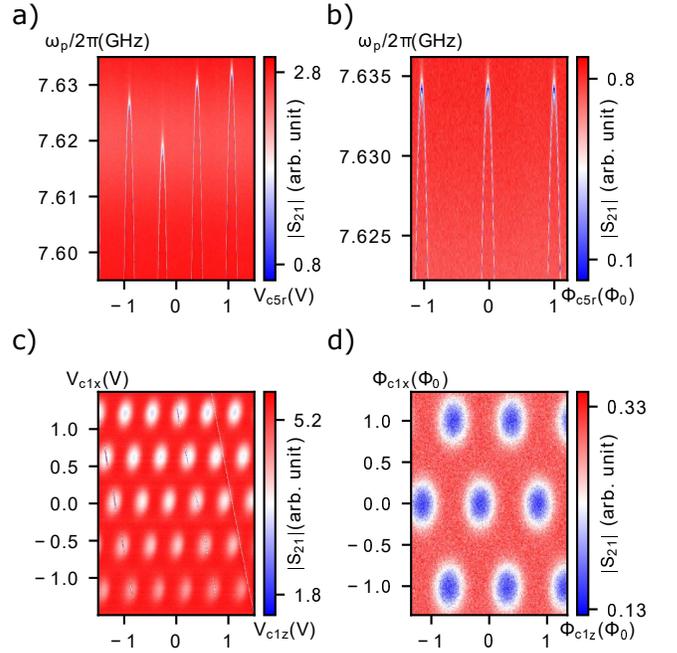}
    \caption{\label{fig:MorePeriodic}(Color Online) (a, b) Resonator transmission  measurement versus probe frequency and resonator bias voltage for unit $c5$ in iteration 1(a) and 2 (b). (c, d) Resonator transmission measurement sweeping $z, x$ biases for unit $c1$ in iteration 1(c) and 2(d). Clearly, the iteration 2 measurement has much better symmetry as compared to iteration 1.}
\end{figure}

Therefore, it is expected that iteration 1 of CISCIQ for device B gives less accurate estimates of the actual coupling coefficients and flux offsets. This is made apparent by simply examining the measured data. As shown in Fig.~\ref{fig:MorePeriodic}(a, b), the Stage 1 measurement for the resonator in unit $c5$ and Stage 3(a) measurement for unit $c1$ are far from the expected periodic behaviour. Fig.~\ref{fig:MorePeriodic} (c) and (d) show the same scan taken during iteration 2. The periodic behavior is restored. 

\subsection{Flux offset drift}\label{sec:OffsetDriftsData}
As noted earlier, the flux offsets drift even when the device is kept cold. It is important to understand the magnitude and timescale over which the flux drifts occur. To perform annealing experiments on the device, the flux offsets need to be stable over a duration that is much longer than any annealing experiment itself. 

\begin{figure}
    \centering
    \includegraphics[]{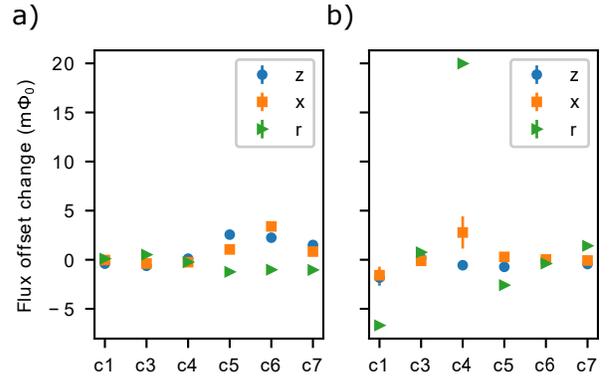}
    \caption{(Color online) \label{fig:OffsetsDrifts}The changes in flux offsets for the $z$ (blue, circle), $x$ (orange, square) and $r$ (green, triangle) loops for each unit. Panel (a) is the change in flux offset after 2 days and panel (b) is after 17 days. The errors are obtained in two steps. First the errors of offsets extracted are computed by first resampling the transmission measurement data with typical measurement noise. The errors in offsets changes are obtained by adding in quadrature the offset errors at different times. Only the coupler units flux offsets are recalibrated in experiments (except $c2$ which had a non-functioning resonator).}
\end{figure}

To check the flux offset stability for device B, after the initial CISCIQi calibration, the flux offsets are recalibrated twice using the method described in Sec.~\ref{sec:OffsetCalibration}. Figure~\ref{fig:OffsetsDrifts} shows the change in flux offsets relative to the initial calibration. After two days, the root mean square (RMS) change in flux offsets for different loops is $1.3\,\text{m}\Phi_0$. After 17 days, one of the resonator SQUID fluxes changed by $20.0\,\text{m}\Phi_0$. The others have an RMS change of $2.0\,\text{m}\Phi_0$. Similar shifts were observed in device A. This suggests that the device can remain well-calibrated for a few days. Over a longer period of time, the flux drifts can be large. Such fluctuations could have various sources, which should be investigated in future work.

\subsection{Characterization of the errors of the calibration protocol}
\begin{figure}
    \centering
    \includegraphics[]{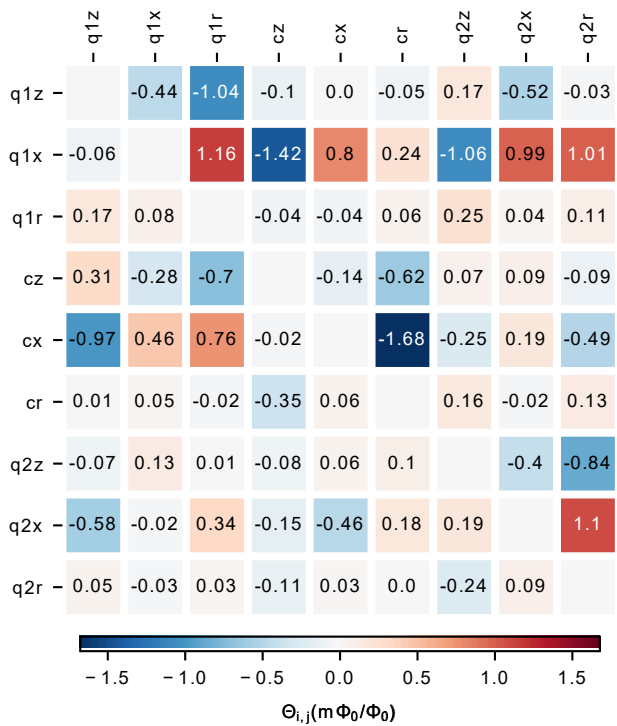}
    \caption{\label{fig:ErrorMatrix}(Color online) The measured crosstalk error coefficients $\Theta_{i,j}$ for each pair of sources (column) and targets (row).}
\end{figure}

The error measurement discussed in Sec.~\ref{sec:ErrorMethod} is applied to device A. The measured errors are displayed in Fig.~\ref{fig:ErrorMatrix}. The RMS of the  errors is $0.5\,\text{m}\Phi_0/\Phi_0$ and the maximum error magnitude is below $1.7\,\text{m}\Phi_0/\Phi_0$. This means that when the estimated flux $f_i^\prime$ is changed by $1\,\Phi_0$ for some loop $i$, while keeping others constant, the actual external flux $f_j$ differs from the approximate coordinate $f_j^\prime$ by at most $1.7\,\text{m}\Phi_0$. In comparison, if no crosstalk compensation is applied, the control error can be lower bounded by the ratio of the final measured crosstalk coefficients $C^{\prime}_{i, j}$ to the direct coupling coefficients $C^\prime_{j,j}$, which has an RMS value of $75\,\text{m}\Phi_0/\Phi_0$. If only one iteration is performed, the errors can be lower bounded by the values of the off-diagonal elements in iteration 2 matrix $C^{(2)\prime}$, which has an RMS value of $2.3\,\text{m}\Phi_0/\Phi_0$ and maximum magnitude of $11\,\text{m}\Phi_0/\Phi_0$. 

The calibration accuracy achieved here is comparable to recent work in Ref.~\cite{abramsMethodsMeasuringMagnetic2019}, where a systematic study of crosstalk calibration was done on a system of superconducting transmon qubits. It is worth mentioning that similar accuracy was achieved in Ref.~\cite{abramsMethodsMeasuringMagnetic2019} using more complex control, involving microwave pulses applied to the qubits. In contrast the method we proposed here only uses resonator transmission measurements.

It is also instructive to compare the calibration error with the quasi-static noise due to low-frequency flux noise intrinsic to the system. Based on flux noise measured in similar devices~\cite{weber_2017_coherentcoupledqubits} and the qubit loop size in our device, the estimated flux noise power spectral density on the qubit $z$-loop is $S_{f_z}(\omega)=A_{f_z}^2/(\omega/2\pi)^\alpha$, with $A_{f_z}=14.4\,\mu\Phi_0/\sqrt{\text{Hz}}$ and $\alpha=0.91$. The noise magnitude is obtained by integrating the power spectral density over a frequency range determined by the experimentally relevant time scales, which is taken to be $\omega/2\pi\in [1~\text{Hz}, 1~\text{GHz}]$. This gives the fluctuation due to flux noise, which is about $281\,\mu\Phi_0$. In comparison, as the maximum variation of flux in any single loop is $1/2\,\Phi_0$, the RMS error due to calibration inaccuracy is $0.5\times 1/2=0.25\,\text{m}\Phi_0$, which is comparable to the intrinsic flux noise.

\subsection{Calibration time}\label{sec:Timescale}
In this subsection we discuss the time taken to complete the calibration protocol. For device A, the first iteration takes about 22 hours while each further iteration takes about 8 hours. Offsets calibration takes about an hour. For device B, the first iteration takes about 80 hours while further iteration takes about 50 hours each. Offsets calibration takes about 2 hours. 

We note that the data acquisition time, which takes about two orders of magnitude more than the data analysis time, is highly specific to the current setup. Firstly, improving the signal to noise ratio could reduce the signal integration time required. This could be done by optimizing the readout frequency and power. Secondly, there is overhead in the software controlling the AWGs. Optimizing the software stack can lead to significant measurement speedup, especially when the number of AWG channels becomes large. Beyond this, improving the measurement protocol by incorporating multiplexed readout could also reduce the measurement time. 

\section{Comparison with targeted mutual inductances}\label{sec:Comparison}
The calibration measurement also provides valuable feedback to circuit design. One important aspect of the design process is to be able to predict the mutual inductances between bias lines and control loops. The measured coupling coefficients can be converted into mutual inductances using the measured resistances along the bias lines in the fridge. This is compared to the mutual inductances extracted by simulating the device with an electromagnetic solver. As the computational resources required for such a simulation scale poorly with the size of the chip, we chose to simulate a single flux cell consisting of a single CSFQ coupled to a resonator SQUID, and their corresponding bias lines in the full two-tier environment (see Appendix~\ref{app:MutualSimulation}). 

Table~\ref{tab:SimulatedMutuals} shows a comparison of the simulated mutual inductances and the measured mutuals on qubit 1 of device A. There is reasonable agreement between the simulated and measured values. Discrepancies could arise due to more complex return current paths through the ground plane, which are not accounted for when simulating only a restricted area of the chip. Given that all the bias lines are connected to the ground plane in the interposer chip, which is facing the qubit, it is not unexpected that the return current effect becomes important. This could be partially mitigated if the connection between bias line and ground is made further away from the control loops. However, this is ultimately limited by the density of control lines and loops in the circuit. In future designs using the three-tier architecture, there will be more flexibility in designing the ground current return path. We expect such an architecture to give better agreement between designs and actual devices.

\begin{table*}
\caption{\label{tab:SimulatedMutuals}
Comparison of simulated and measured (in brackets) values of mutual inductances between bias lines and loops.}
\begin{ruledtabular}
\begin{tabular}{c|c|c|c}
&$z$ bias line &$x$ bias line &$r$ bias line\\
\hline
Qubit Loop $z^\prime$ \footnote{Due to the specific convention used, $z$-loop does not refer to a physical loop in the device, therefore $z^\prime$-loop is used instead (see Appendix \ref{app:FluxConvention}).}& $-1.6 (-1.9)$ pH & $-0.65(-1.3)$ pH & $-0.21(-0.2)$ pH \\
\hline
Qubit Loop $x$& $0.0098(-0.065)$ pH & $1.4(2.5)$ pH & $0.0003(-0.0068)$ pH\\
\hline
Resonator Loop& $0.25(0.31)$ pH & $-0.021(-0.028)$ pH & $-1.4(-1.7)$ pH\\
\end{tabular}
\end{ruledtabular}
\end{table*}

\section{Summary and Conclusion}\label{sec:Summary}
In summary, we proposed and implemented an iterative approach to calibrate flux crosstalk, which only relies on the symmetry properties of superconducting circuits, without needing a full model of the device. The efficacy of the flux crosstalk calibration is clearly validated by the convergence of the crosstalk and offsets measured in each iteration, as well as the independent error measurement. The iterations address the errors due to strong inductive coupling within the circuit. Detailed experiments were performed on two devices, labeled A and B. Comparing the calibration results between device A and B, it is clear that device B, which has more couplers, requires more iterations to achieve the same level of convergence. This highlights the importance of iteration when calibrating devices with strong inter-element interaction. 

In terms of flux control complexity these devices are comparable to some of the largest gate-based flux tunable circuits. When considering applying the calibration method in this work to future large-scale quantum processors, we can look at the problem from two different perspectives. First from the design and fabrication perspective, future generation devices are likely to incorporate multi-tier architectures such as the ones in Ref.~\cite{yostSolidstateQubitsIntegrated2020}. Such architecture allows current signals to be routed away from the circuit loops before they are grounded. With this advance we expect the crosstalk to be more spatially localized, so that the number of crosstalk elements to be measured should only scale as $N$, instead of $N^2$. This however, would not eliminate the need for iterations, which addresses the issue of strong inter-element interaction. Secondly, from a measurement perspective, we expect further technical developments to speed up the data acquisition, as discussed in Sec.~\ref{sec:Timescale}.

The crosstalk calibration method in this work was applied to devices developed specifically for quantum annealing applications. Compared to commercial quantum annealers~\cite{johnson_2010_scalablecontrolsystem, bunyk_2014_architecturalconsiderationsdesign, kingPerformanceBenefitsIncreased2020}, we explore an implementation with independent local high-bandwidth control of qubits and couplers, enabling advanced annealing protocols, and simplified circuits without built-in compensation for variation in fabrication parameters, leading to increased coherence. While creating new opportunities for quantum annealing, this design approach leads to increased complexity of flux crosstalk calibration, a challenge that can be tackled with the methods we developed here.

\begin{acknowledgments}
We thank the members of the Quantum Enhanced Optimization/Quantum Annealing Feasibility Study collaboration for various contributions that impacted this research. In particular, we thank K. Zick and D. Ferguson for fruitful discussion of experiments, A. J. Kerman for the guidance on circuit simulations and design, and R. Yang and S. Bedkihal for related work on circuit modelling and useful discussions. We gratefully acknowledge the MIT Lincoln Laboratory design, fabrication, packaging, and testing personnel for valuable technical assistance. The research is based upon work supported by the Office of the Director of National Intelligence (ODNI), Intelligence Advanced Research Projects Activity (IARPA) and the Defense Advanced Research Projects Agency (DARPA), via the U.S. Army Research Office contract W911NF-17-C-0050. The MIT Lincoln Laboratory work was funded under Air Force Contract No. FA8702-15-D-0001. The views and conclusions contained herein are those of the authors and should not be interpreted as necessarily representing the official policies or endorsements, either expressed or implied, of the ODNI, IARPA, DARPA, or the U.S. Government. The U.S. Government is authorized to reproduce and distribute reprints for Governmental purposes notwithstanding any copyright annotation thereon.
\end{acknowledgments}
\newpage

\appendix


\section{Experimental setup}\label{app:FridgeWiring}
\begin{figure*}
\centering
\includegraphics[]{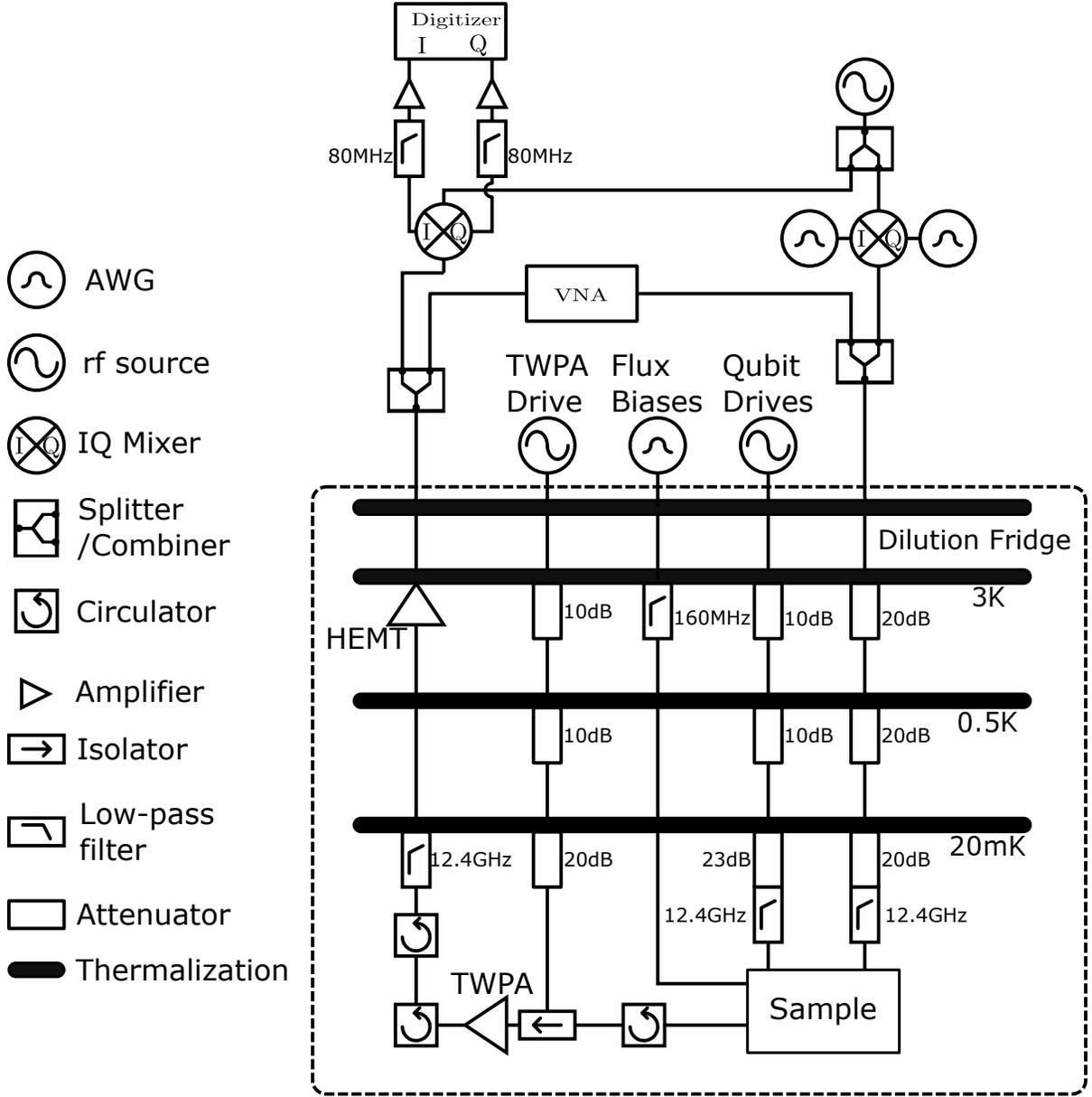}
\caption{\label{fig:FridgeWiring}Schematic of the wiring setup inside and outside of the dilution fridge (see text for additional explanations).}
\end{figure*}

In Fig.~\ref{fig:FridgeWiring} we present a diagram of the setup showing the room temperature electronics and the wiring inside the dilution refrigerator. Arbitrary waveform generators (AWG) are used for flux biasing the circuit. The AWGs (Keysight AWG3202A) can supply voltages ranging from $-1.5\,\text{V}$ to $+1.5\,\text{V}$ with 14 bits of precision. The bias current is carried to the packaged chip through twisted wires. Inline resistors of appropriate resistances (typically $1000\,\Omega$) are used to apply current leading to a voltage to flux conversion of the order of $1\,\Phi_0/V$. The twisted wires have a limited bandwidth of about $10\,\text{MHz}$. Later generations of the experiment employ customized wiring, which are designed to have higher bandwidth.

A heterodyne readout setup is used for probing the resonator readout. A  readout pulse is generated starting with a stable tone generated by a microwave synthesizer, which is then upconverted using an IQ mixer. At the output of the device, the output signal is first amplified by a travelling-wave-parametric-amplifier (TWPA)~\cite{macklin_2015_quantumlimitedjosephsontravelingwave} at the mixing chamber of the dilution refrigerator and by a high-electron mobility transistor (HEMT) amplifier at the 3K stage of the dilution refrigerator. It then gets downconverted at room temperature before its quadrature voltages are sampled using a digitizer.

\section{Qubit and coupler loop geometry}\label{app:FluxConvention}

\begin{figure}[t]
    \centering
    \includegraphics[]{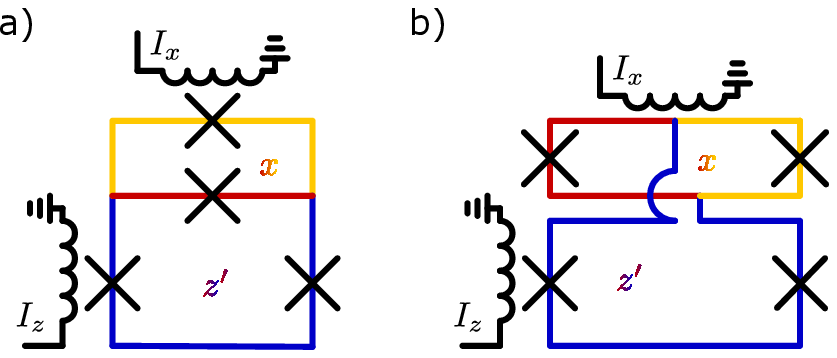}
    \caption{(Color online) (a)A conventional 4 junction qubit geometry. (b) The symmetrized x-loop qubit geometry. In both cases the blue and red arms form the $z^\prime$-loop in the text and the yellow and red arms form the $x$-loop.}
    \label{fig:LoopGeometry}
\end{figure}

The qubits and the couplers have two fundamental loops, denoted by $z^\prime, x$  in Fig.~\ref{fig:LoopGeometry}(a). One can define two independent external fluxes $f_{z^\prime}$ and $f_{x}$, as shown in Fig.~\ref{fig:LoopGeometry}. As demonstrated in one of the earliest proposals of flux qubits \cite{mooij_1999_josephsonpersistentcurrentqubit}, $f_z= f_{z^\prime}+1/2f_x=(1/2+n)$ corresponds to the symmetry point of the circuit, provided the two junctions in the $x$-loop have identical sizes \cite{khezriAnnealpathCorrectionFlux2021}. Around the symmetry point, the Hamiltonian projected onto the subspace formed by the lowest two energy eigenstates can be written as
\begin{align}
    H = -I_p(f_x)\left(f_z-\frac{1}{2}\right)\Phi_0\sigma_z - \frac{\Delta(f_x)}{2}\sigma_x,
\end{align}
where $I_p$ is the persistent current and $\Delta$ is the gap at the symmetry point. Note that both of these quantities depend on $f_x$. During a typical quantum annealing experiment, one starts with large $\Delta$ and ends with minimal $\Delta\approx0$. This corresponds to varying $f_x$ from $1/2$ to $1$, while keeping $f_z$ close to the symmetry point. As the flux qubit energy is very sensitive to $f_z$, it is beneficial to keep the excursion in its corresponding bias current $I_z$ small during the annealing, so that pulse distortion does not cause significant errors. This motivates the symmetrized $x$-loop design, as used in the coupler design in Ref.~\cite{ harris_2009_compoundjosephsonjunctioncoupler}. An illustration of the design is shown in Fig.~\ref{fig:LoopGeometry}(b). In this geometry, the mutual between the $x$ bias line the and $z^\prime$-loop, $M_{z^\prime x}$, is designed to be about $-1/2$ of the mutual between the $x$ bias line and the $x$-loop, $M_{xx}$. The relation between the two control loop fluxes and bias current becomes
\begin{align}
    \begin{pmatrix}
        f_z\\
        f_x\\
    \end{pmatrix}
    &=\begin{pmatrix}
        1&\frac{1}{2}\\
        0&1
    \end{pmatrix}
    \begin{pmatrix}
        f_{z^\prime}\\
        f_x
    \end{pmatrix}\\
    &=\begin{pmatrix}
        1&\frac{1}{2}\\
        0&1
    \end{pmatrix}
    \begin{pmatrix}
        M_{z^\prime z}&&M_{z^\prime x}\\
        M_{x z}&&M_{x x}\\
    \end{pmatrix}
    \begin{pmatrix}
        I_z\\
        I_x
    \end{pmatrix}\\
    &=\begin{pmatrix}
        M_{z^\prime z}+\frac{1}{2}M_{x z}&M_{z^\prime x}+\frac{1}{2}M_{x x}\\
        M_{x z}&M_{x x}
    \end{pmatrix}
    \begin{pmatrix}
        I_z\\
        I_x
    \end{pmatrix}.
\end{align}
Since $M_{z^\prime x}+\frac{1}{2}M_{x x}\approx0$ by design, $f_z$ remains largely unchanged when $I_x$ pulses are applied. This hence minimizes the excursions in $I_z$ during annealing experiments. 

Since $z^\prime$,$x$ loops are physical loops in the superconducting circuit, properties of the circuit can be considered as a lattice in the $f_{z^\prime}, f_{x}$ coordinates, with primitive lattice vectors $(1, 0)^T$ and $(0, 1)^T$. In the flux coordinates $f_z, f_x$, the primitive lattice vectors become $(1, 0)^T$ and $(1/2, 1)^T$. During CISCIQ iterations and error measurements, it is easier to work with another set of independent lattice vectors $(1, 0)^T$ and $(0, 2)^T$ in the $f_z, f_x$ coordinates. This means that when periodic steps are used for the crosstalk source bias, the choice of step for $x$ bias becomes $2\,\Phi_0$  instead of $1\,\Phi_0$.

\section{Symmetries in external fluxes in CSFQs and couplers}\label{app:FluxSymmetry}
In Stage 3(a) of CISCIQ, we use the fact that the CSFQs and the couplers exhibit point reflection symmetries with respect to half integer flux quanta points in both the $x$ and $z^\prime$ loops. This property is derived from two underlying symmetries in the circuit. First, a single flux cell has mirror symmetry about the chip plane, so that the resonator frequency should have $\omega_r(f_{z^\prime}, f_x, f_r)=\omega_r(-f_{z^\prime}, -f_x, -f_r)$. Second, superconducting loops have properties periodic in $\Phi_0$, so that $\omega_r(f_{z\prime}, f_x, f_r)=\omega_r(f_{z^\prime}+1, f_x+1, f_r+1)$. Combining these two relations we find that $\omega_r(f_{z^\prime}+N_z/2, f_x+N_x/2, f_r+N_r/2)=\omega_r(-f_{z^\prime}+N_z/2, -f_x+N_x/2, -f_r+N_r/2)$, where $N_{(z^\prime,x,r)}$ are integers. 

In Stage 3 we assumed that each flux cell is isolated from the rest, and the resonator calibration is exact, so that $f_r=0$. The fluxes $f_{z^\prime}$ and $f_x$ are completely specified by the affine transformation (see Sec.~\ref{sec:CISCIQ}).

\begin{align}
        \begin{pmatrix}
        f_{z^\prime} \\
        f_x
        \end{pmatrix}
        &=
        \begin{pmatrix}
        1 & -\frac{1}{2}\\
        0 & 1    
        \end{pmatrix}
        \begin{pmatrix}
        f_z\\
        f_x\\
        \end{pmatrix}\\
        &=
        \left(
        \begin{array}{cc}
        C_{z,z}^{\text{eff}} & C_{z,x}^{\text{eff}}-\frac{1}{2}C_{x,z}^{\text{eff}}\\
        C_{x,z}^{\text{eff}} & C_{x,x}^{\text{eff}}\\
        \end{array}
        \right)
        \begin{pmatrix}
        V_z \\
        V_x
        \end{pmatrix}\nonumber\\
        &+\begin{pmatrix}
        f_{0,z}^{\text{eff}}-\frac{1}{2}f_{0,x}^{\text{eff}}\\
        f_{0,x}^{\text{eff}}\\
        \end{pmatrix}.
\end{align}
The analysis for Stage 3(a) data relies on the fact that the data has point reflection symmetry in the $V_z, V_x$ coordinates. Hence, we need to show that point reflection symmetries are preserved under affine transformation. To show this, consider two sets of points $A$ and $B$, related by an affine transformation $\mathcal{F}$, such that $\mathcal{F}(B)=A$. It is also known that $A$ has point reflection symmetry so that $\mathcal{R}(A)=A$. Here $\mathcal{R}$ is the point reflection operation with the property $\mathcal{R}=\mathcal{R}^{-1}$. To show that $B$ also possesses some point reflection symmetry, we need to a) find some operation $\mathcal{R}^\prime$ which satisfies the symmetry condition $\mathcal{R}^\prime(B)=B$ and b) show that $\mathcal{R}^\prime$ is indeed a point reflection operation. To find $\mathcal{R}^\prime$ we write
\begin{align}
    &\mathcal{F}(B)=A=\mathcal{R}(A) = \mathcal{R}\circ\mathcal{F}(B),\\
    &\mathcal{F}^{-1}\circ\mathcal{R}\circ\mathcal{F}(B)=B.
\end{align}
Hence $\mathcal{R}^\prime\equiv\mathcal{F}^{-1}\circ\mathcal{R}\circ\mathcal{F}$ satisfies the symmetry condition. To show that $\mathcal{R}^\prime$ is indeed a point reflection operation, we just need to show that it is isometric and involutive. It is involutive because we can write
\begin{align}
    \mathcal{F}^{-1}\circ\mathcal{R}\circ\mathcal{F}\circ\mathcal{F}^{-1}\circ\mathcal{R}\circ\mathcal{F}&=\mathcal{I},
\end{align}
where $\mathcal{I}$ is the identify operation and we used $\mathcal{R}\circ\mathcal{R}=\mathcal{I}$. It can be shown that $\mathcal{R}^\prime$ is also isometric because it is composed of $\mathcal{F}^{-1}$, $\mathcal{R}$ and $\mathcal{F}$, which are individually isometric. Therefore point reflection symmetry holds both in the flux coordinates $f_z^\prime, f_x$ as well as the voltage coordinates $V_z, V_x$.

\section{Recurrence analysis and line detection}\label{app:RecurrenceAnalysis}

This section discusses the various image processing applied during the analysis of Stage 1 of CISCIQ. In this stage, resonator transmission is measured as a function of probe frequency and resonator bias. As the background transmission is different at different frequencies, a background filter is applied, which is specified by,
\begin{align}
    S_{21}^\prime(\omega_p, V_r)=\frac{S_{21}(\omega_p, V_r) }{\mathbb{Md}\left(S_{21}(\omega_p)\right)},
\end{align}
where $S_{21}^\prime$ stand for the filtered results and $\mathbb{M}$ stand for taking the complex median over the $V_r$ dimension.

\begin{figure}
    \centering
    \includegraphics[]{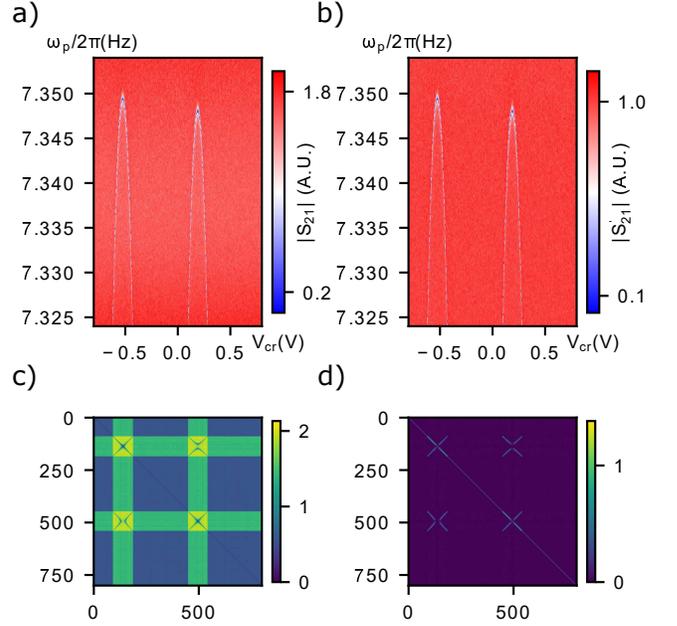}
    \caption{(Color online) Top row shows the resonator S21 measurement before (left) and after (right) the image processing routines. Bottom left plot shows the corresponding column pair-wise distance for the top right image. The recurrence plot (bottom right) for the top right image is obtained by thresholding the column pair-wise distances image.}
    \label{fig:ResonatorDataCleaning}
\end{figure}

Furthermore, to enhance the resonance dip feature relative to the background, a median filter is applied to the image along the frequency axis. The raw measurement data and the data after applying the background and median filters are shown in Fig.~\ref{fig:ResonatorDataCleaning}(a) and (b).

To obtain the recurrence plot, the first step is to compute the pair-wise distance between columns of the image. This calculation results in Fig.~\ref{fig:ResonatorDataCleaning}(c). Then the pair-wise distance image goes through Sobel horizontal and vertical filters sequentially to enhance the features that correspond to translational symmetry. This results in Fig.~\ref{fig:ResonatorDataCleaning}(d). Finally, the filtered image is thresholded using Otsu thresholding ~\cite{otsuThresholdSelectionMethod1979} to give the recurrence plot, which is Fig.~\ref{fig:RecurrenceAnalysis}(d) in the main text. To identify lines and thus translations, the Hough transform is applied. This then completes our custom implementation of translational symmetry detection. Compared to readily available image registration functions, the custom algorithm allows specifying ranges within which to look for translations, hence avoiding finding translations that are multiple periods away. 

\section{Coupler-resonator inductive loading model}\label{app:CouplerResonator}
This section describes in detail the inductive loading model between a single coupler and its coupled tunable resonator used in Sec.~\ref{sec:CISCIQ}. We start by defining the effective quantum inductance of the coupler, $L^{\text{eff}}_{C}$, based on~\cite{weber_2017_coherentcoupledqubits},
\begin{align}
    \frac{1}{L^{\text{eff}}_{C}} = \frac{1}{\Phi_0}\frac{\partial \langle I^C\rangle }{\partial f_{cz}},
\end{align}
where $L^{\text{eff}}_C$ is the coupler effective inductance, $\langle I^C \rangle$ is the ground state current in the coupler $z$-loop and $f_{cz}$ is the $z$-flux bias for the coupler. The quantity $L^{\text{eff}}_{C}$ is obtained using the full circuit model of the coupler~\cite{Daniel2021} and a quantum circuit simulation package~\cite{kermanEfficientNumericalSimulation2020}.

The tunable resonator can be modeled as a waveguide terminated to ground through the effective inductance of the rf-SQUID. For a classical rf-SQUID with junction critical current $I_c$ and geometric inductance $L_{g}$, its effective inductance is given by
\begin{align}
    \frac{1}{L_{\text{SQUID}}^{\text{eff}}} = \frac{1}{L_g} + \frac{2\pi I_c \cos\varphi}{\Phi_0},
\end{align}
where $\varphi$ is the phase across the junction. The phase $\varphi$ can be found by minimizing the SQUID classical potential
\begin{align}
    U(\varphi)=-\frac{I_c\Phi_0}{2\pi} \cos(\varphi)+\frac{\Phi_0^2}{2L_g}\l(\frac{\varphi}{2\pi}-f_r\r)^2.
\end{align}
Then the resonance frequency $\omega_r$ for the $\lambda/4$ waveguide together with the rf-SQUID is found by numerically solving the equation
\begin{align}
    \exp\l(\frac{2i\omega_r l}{c}\r)=\frac{i\omega_r L^{\text{eff}}_{\text{SQUID}}-Z_0}{i\omega_r L^{\text{eff}}_{\text{SQUID}}+Z_0},
\end{align}
where $l, c, Z_0$ are the waveguide length, phase velocity and characteristic impedance respectively.

With inductive loading, the geometric inductance of the SQUID changes via
\begin{align}
    L_g \xrightarrow[]{} L_{g} - \frac{M_{\text{coupler},\text{SQUID}}^2}{L^{\text{eff}}_C},
\end{align}
where $M_{\text{coupler},\text{SQUID}}$ is the mutual inductance between coupler $z$ and rf-SQUID loops. As the coupler bias changes, its effective inductance also changes, which then changes the SQUID effective inductance and resonator frequency. 
\section{Coupling matrices and offsets for each iteration}\label{app:IterationResults}
\begin{figure*}
    \centering
    \includegraphics[]{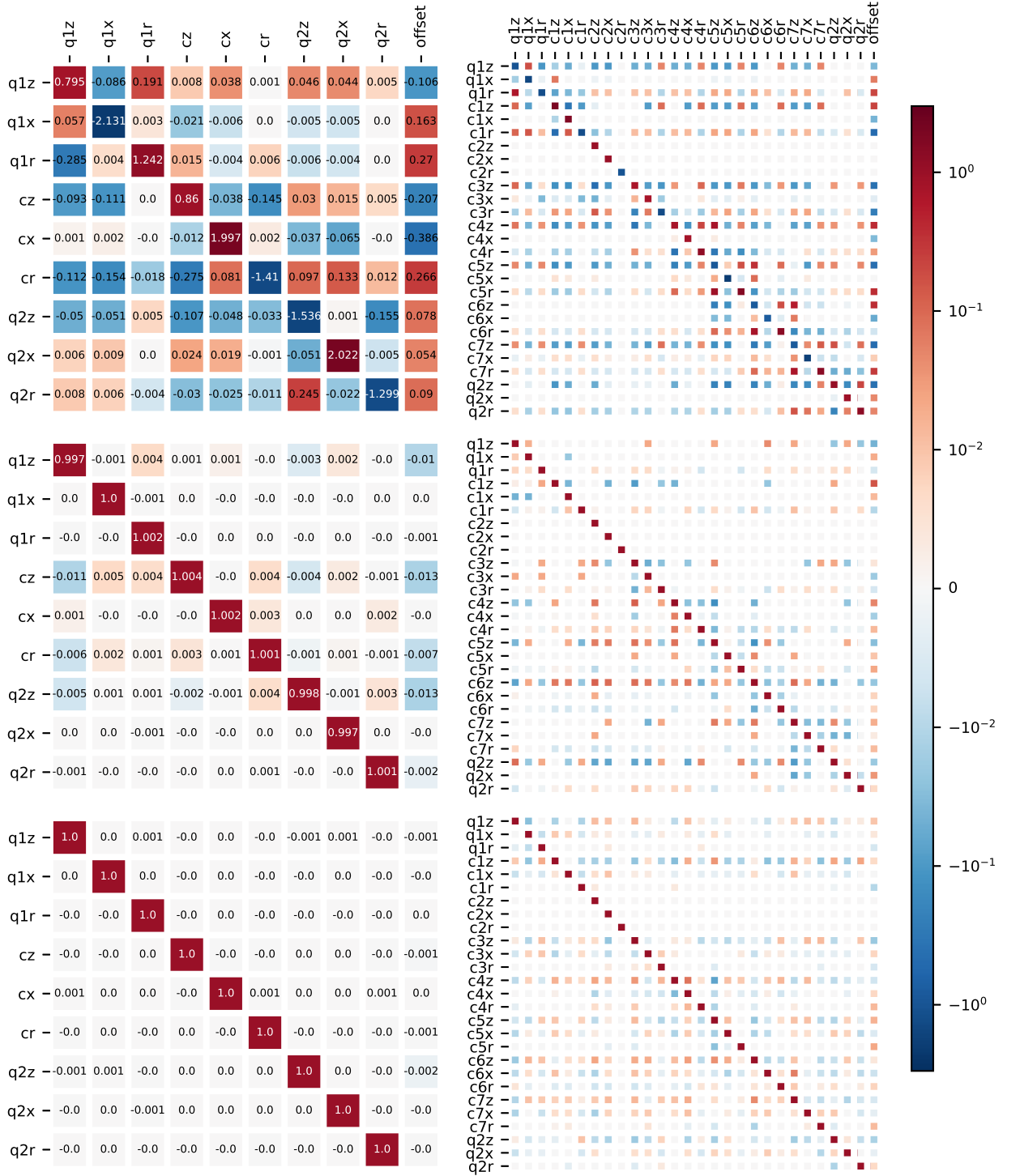}
    \caption{(Color online) Left (right) three panels show for device A (B) the coupling matrices and flux offsets measured for iteration 1 (top), 2 (middle) and 3 (bottom)}
    \label{fig:IterationMatrices}
\end{figure*}

Figure~\ref{fig:IterationMatrices} presents coupling matrices $\mathbf{M}^{(n)\prime}$ and flux offsets $\mathbf{f}_0^{(n)\prime}$ measured at each iteration of CISCIQ for both devices A and B. As can be seen from the iteration 2 and 3 results, for both devices, the convergence is indicated by the decreased intensity of the colors on the off-diagonal elements and flux offsets, as well as the diagonal elements approaching 1.

\section{Circuit parameters}\label{app:CircuitParameters}
To give more concrete numbers on the strength of circuit interaction, Tab.\ref{tab:IpsMutuals} tabulates the range of persistent currents for the qubits, couplers and resonator SQUIDs, as well as their geometric mutual inductances. For the qubit and couplers, the persistent currents are found by numerically solving the quantum circuit Hamiltonian. For the resonator, the current is calculated by solving the classical rf-SQUID equation. A complete list of the circuit parameters will be presented in a related paper ~\cite{Daniel2021}. 
\begin{table}[b]
\caption{Circuit parameters and their corresponding values. \label{tab:IpsMutuals}}
    \begin{ruledtabular}
    \begin{tabular}{ll}
         Circuit parameter & Value\\
         \hline
         $I_p$(qubit) & within $\pm0.14\mu\text{A}$\\
         $\langle I^{C}\rangle$(coupler) & within $\pm0.45\mu\text{A}$\\
         $I_p$(SQUID) & within $\pm1.2\mu\text{A}$\\
         $M_{\text{qubit}, \text{SQUID}}$&  29.5($\text{m}\Phi_0/\mu\text{A}$)\\
         $M_{\text{coupler},\text{SQUID}}$&
         28.7($\text{m}\Phi_0/\mu\text{A}$)\\
         $M_{\text{qubit}, \text{coupler}}$&
         30.2($\text{m}\Phi_0/\mu\text{A}$)\\
         $M_{\text{coupler}, \text{coupler}}$& 31.0($\text{m}\Phi_0/\mu\text{A}$)\\
    \end{tabular}
    \end{ruledtabular}
\end{table}

The maximum possible induced fluxes from one circuit element to another is $36\,\text{m}\Phi_0$. This is consistent with the fact that about $10\,\text{m}\Phi_0$ of error is measured on device B when only one iteration of CISCIQ is applied. 

\section{Simulation of mutual inductances using Sonnet}\label{app:MutualSimulation}
To simulate the mutual inductances, the design drawings are first imported into Sonnet, a microwave modelling software for 3D planar circuits. The model includes both the interposer and qubit layer, as well as all the bump bonds and air bridges. Gaps in the superconducting loop left for Josephson junctions are connected in the simulation. Ports are assigned to each superconducting loop and bias lines. The inductances are extracted by computing the impedance matrix at $1\,\text{GHz}$. It is also found that there is little dependence on frequency. 

\section{Error due to measurement noise}\label{app:MeasurementNoiseError}
\begin{figure}
    \centering
    \includegraphics[]{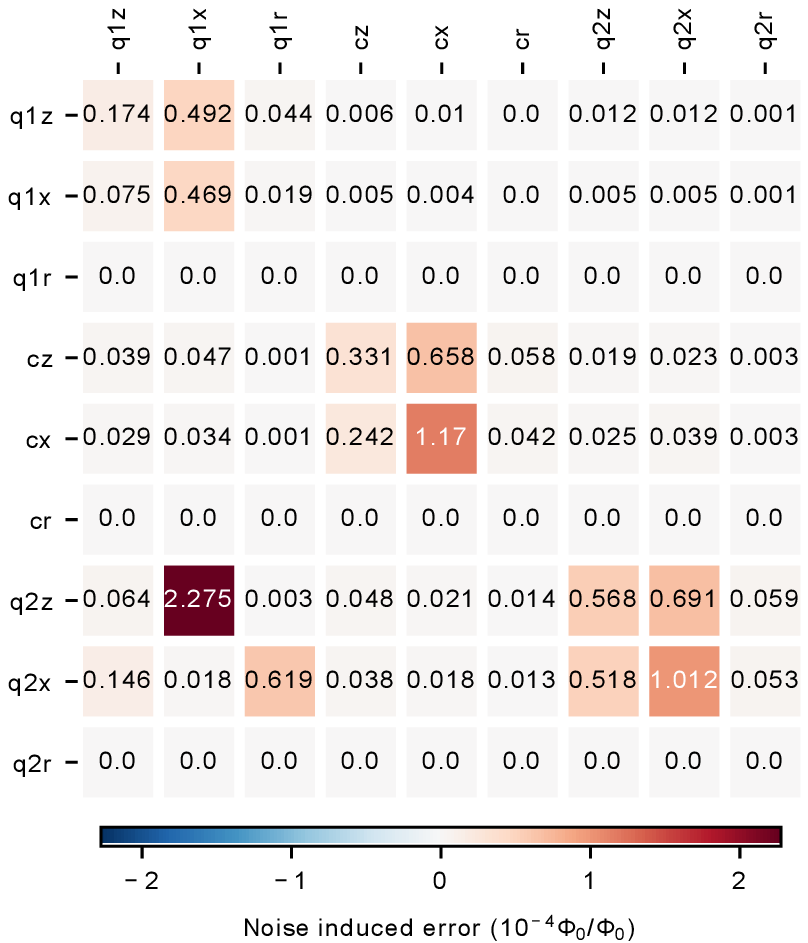}
    \caption{(Color online) Measurement noise induced error in the coupling matrix for device B. The induced errors on the measured crosstalk to resonators are much smaller than to the qubits.}
    \label{fig:MeasurementNoise}
\end{figure}

Since the analysis of CISCIQ data relies heavily on identifying symmetries in the measured $S_{21}$ images, one could ask whether the fluctuations in $S_{21}$ due to measurement noise causes significant error. For this reason we characterized the error of the coupling coefficients solely due to measurement noise. This is done by resampling the measurement data with added Gaussian noise on measured $|S_{21}|$. The noise parameters are chosen to reflect typical values at the choice of measurement parameters, such as the number of repetitions and readout integration time. We apply re-sampling on the data taken during the last iteration of device B. After applying the analysis procedure on 100 sets of resampled data, the standard deviation of the resultant coupling matrix is plotted in Fig.~\ref{fig:MeasurementNoise}. The largest element is $0.2\text{m}\Phi_0/\Phi_0$, about 10 times lower than the total error measured in the main text. This shows that the error of calibration is not limited by the measurement noise.

\bibliography{references}

\end{document}